\documentclass[]{elsart5p}

\usepackage{graphicx}
\usepackage{rotating}
\usepackage{amssymb}
\usepackage[english]{babel}

\begin{document}

\begin{frontmatter}



\title{TRIGA-SPEC: A setup for mass spectrometry and laser spectroscopy at the research reactor TRIGA Mainz}


\author[1]{J. Ketelaer\corauthref{cor1}\ead{ketela@uni-mainz.de}},
\author[2]{J. Kr\"{a}mer},
\author[3]{D. Beck},
\author[1,3,4]{K. Blaum},
\author[3]{M. Block},
\author[2]{K. Eberhardt},
\author[1]{G. Eitel},
\author[1]{R. Ferrer\thanksref{msu}},
\author[2,3]{C. Geppert},
\author[1,3]{S. George},
\author[3]{F. Herfurth},
\author[1]{J. Ketter},
\author[1]{Sz. Nagy},
\author[1]{D. Neidherr},
\author[2]{R. Neugart},
\author[2,3]{W. N\"{o}rtersh\"{a}user},
\author[1]{J. Repp},
\author[1]{C. Smorra},
\author[2]{N. Trautmann},
\author[5]{C. Weber}.

\corauth[cor1]{Corresponding author. Johannes Gutenberg-Universit\"{a}t Mainz, Institut f\"{u}r Physik, Staudinger Weg 7, D-55128 Mainz.\\Tel.: +49 6131 39 24151; Fax: +49 6131 39 23428.}

\thanks[msu]{Present address: National Superconducting Cyclotron Laboratory, MSU, East Lansing, MI 48824-1321, USA.}

\address[1]{Institut f\"{u}r Physik, Johannes Gutenberg-Universit\"{a}t Mainz, Staudinger Weg 7, D-55128 Mainz, Germany.}
\address[2]{Institut f\"{u}r Kernchemie, Johannes Gutenberg-Universit\"{a}t Mainz, Fritz-Stra\ss mann-Weg 2, D-55128 Mainz, Germany.}
\address[3]{Gesellschaft f\"{u}r Schwerionenforschung mbH, Planckstra\ss e 1, D-64291 Darmstadt, Germany.}
\address[4]{Max-Planck-Institut f\"ur Kernphysik, Saupfercheckweg 1, D-69117 Heidelberg, Germany.}
\address[5]{University of Jyv\"{a}skyl\"{a}, P.O. Box 35 (YFL), FI-40014 Jyv\"{a}skyl\"{a}, Finland.}

\begin{abstract}
The research reactor \textsc{Triga} Mainz is an ideal facility to provide neutron-rich nuclides with production rates sufficiently large for mass spectrometric and laser spectroscopic studies. Within the \textsc{Triga-Spec} project, a  Penning trap as well as a beam line for collinear laser spectroscopy are being installed. Several new developments will ensure high sensitivity of the trap setup enabling mass measurements even on a single ion. Besides neutron-rich fission products produced in the reactor, also heavy nuclides such as \textsuperscript{235}U or \textsuperscript{252}Cf can be investigated for the first time with an off-line ion source. The data provided by the mass measurements will be of interest for astrophysical calculations on the rapid neutron-capture process as well as for tests of mass models in the heavy-mass region. The laser spectroscopic measurements will yield model-independent information on nuclear ground-state properties such as nuclear moments and charge radii of neutron-rich nuclei of refractory elements far from stability. This publication describes the experimental setup as well as its present status.
\end{abstract}

\begin{keyword}
Penning trap \sep mass spectrometry \sep laser spectroscopy \sep
heavy nuclides \sep fission products \sep nuclear reactor
\PACS 07.75.+h mass spectrometers \sep 21.10.Dr binding energies and
masses \sep 32.10.Fn fine and hyperfine structure \sep 31.30.Gs
hyperfine interactions and isotope effects
\end{keyword}

\end{frontmatter}


\section{Introduction and motivation}

\label{sec_introduction} Our world is determined to a large degree by the properties of nuclear matter, which can be studied even far away from the valley of stability at the presently existing large-scale radioactive beam facilities. While only light nuclei of hydrogen, helium and lithium were created in the big bang, all other elements emanate from nuclear fusion in stars. However, the nuclear binding energy has a maximum at \textsuperscript{56}Fe, which leads to the question, why heavier elements exist in nature at all, since there is no further energy gain in fusion reactions \cite{scha2006}. In astrophysics, proton- and neutron-capture processes are assumed to create elements heavier than iron \cite{burb1957,arno2003,cowa2006}. On the neutron-rich side of the nuclear chart, nuclides are produced by a neutron-capture series prior to $\beta^-$-decay. Two different processes have to be distinguished here: if the neutron flux is large enough to go far from the valley of stability, the element synthesis follows the path of the r-process, while the path close to the stable nuclides is called s-process. Considering the r-process on the neutron-rich side of the nuclear chart, an important input parameter for reliable calculations of the abundances of isotopes along its path are mass values reflecting the nucleonic interactions. The neutron-capture process will cease at a point where the binding energy for the next additional neutron is such low that an equilibrium between the reactions (n,$\gamma$) and ($\gamma$,n) is reached \cite{burb1957}. The r-process is waiting until a $\beta^-$-decay occurs, which enables further capture events. Hence, ground-state properties of nuclei, such as masses, lifetimes and decay modes, significantly determine the paths of element formation. Thus, high-precision measurements of masses, spins and moments, provide important ingredients for reliable nuclear astrophysical calculations and for the understanding of the composition of matter in the Universe in general. Research in this direction has been pursued by many Penning trap mass spectrometric and laser spectroscopic setups installed at radioactive beam facilities world-wide \cite{blau2006}.

So far, the region close to the r-process path is rather unexplored due to low production rates of the involved nuclei at all presently existing radioactive beam facilities and thus our knowledge on the ground-state properties of almost half of the heavy nuclides is scarce. In the coming future this may change dramatically since large radioactive beam facilities currently being proposed, such as \textsc{Fair} (Darmstadt, Germany) \cite{fair}, \textsc{Spiral 2} (Caen, France) \cite{spir}, \textsc{Eurisol} \cite{euri}, and \textsc{Ria} (USA) \cite{ria}, will extend the accessible part of the nuclear chart significantly. Within this article we describe the installation of a mass spectrometric and laser
spectroscopic setup, called \textsc{Triga-Spec}, at the research reactor \textsc{Triga} Mainz \cite{eber2000,hamp2006}, which gives us already prior to the start of these large future facilities improved access to the neutron-rich nuclei in the region of the r-process and allows us to perform high-precision measurements of ground-state properties in this region.

\begin{figure*}[tbp]
\begin{center}
\includegraphics* [width=0.7\textwidth]{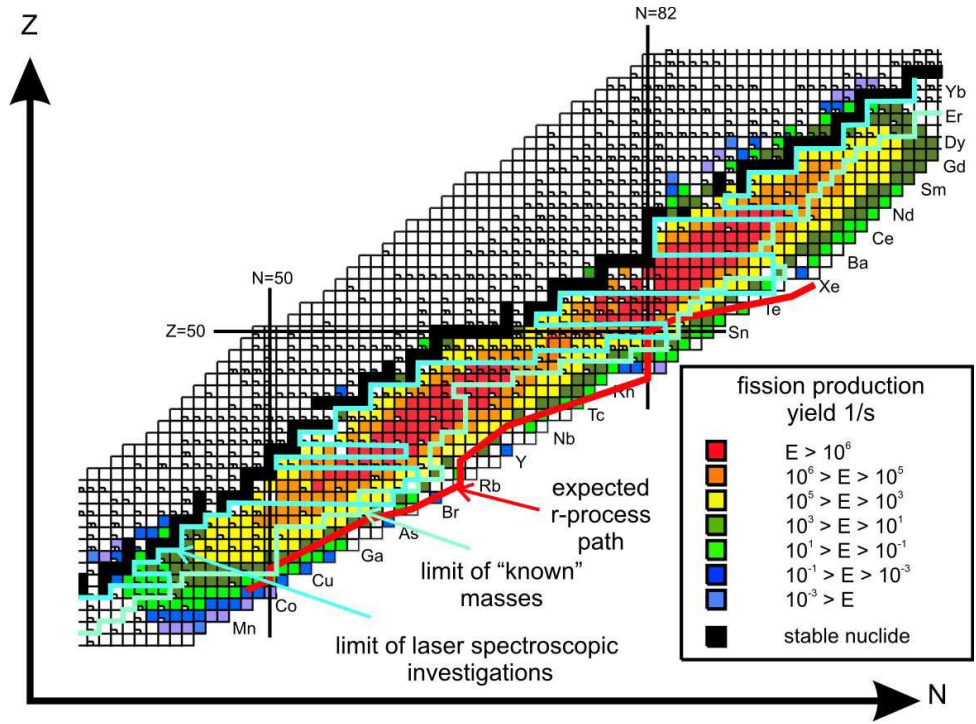}
\end{center}
\caption{(Color) Production rates of fission nuclides with a $300\,\mu$g \textsuperscript{249}Cf target and a thermal neutron flux of $1.8\cdot 10^{11}\,$n/(cm\textsuperscript{2}s) at the \textsc{Triga} Mainz reactor as obtained in the steady-state operation mode. For details see text.}
\label{fig_yields}
\end{figure*}

Figure \ref{fig_yields} gives an example of the expected production rates at the \textsc{Triga} Mainz for a \textsuperscript{249}Cf target. The rates have been calculated assuming a $300\,\mu$g target and a thermal neutron flux of $1.8\cdot 10^{11}\,$n/(cm$^{2}$s), which is typically available at this research reactor in the continuous operation mode. The yields for thermal-neutron induced fission are taken from \cite{berk}. Particularly interesting are the isotopes around the magic neutron number $N=82$, since the r-process approaches known nuclides in this region. Some of the exotic neutron-rich species are produced at comparably low rates in the order of a few particles per second, but they have rather long half-lives in the order of seconds. Therefore, these are ideal candidates for mass spectrometry with the newly developed non-destructive narrow-band image current detection system (\textsc{Ft-Icr}: Fourier Transform-Ion Cyclotron Resonance), which will pave the way to perform a measurement with a single stored ion in a trap. For a number of these isotopes, this will be the first direct mass measurement. \textsc{Triga-Spec} is presently the only facility to study nuclear ground-state properties at a research reactor. Similar installations or similar experiments have been proposed at the \textsc{Frm-II} reactor in Garching \cite{habs2006} and within the \textsc{Caribu} project (Argonne National Lab) \cite{sava2006}. 

The \textsc{Triga} reactor, located at the Institute for Nuclear Chemistry of the University of Mainz, has a steady-state thermal power of 100\,kW \cite{hamp2006}. It can also be operated in a pulsed mode with a repetition rate of one pulse per five minutes and a maximum peak power of 250\,MW. A thin target of a fissionable nuclide, e.g.\,\textsuperscript{249}Cf, can be placed close to the reactor core. The neutron induced fission products are transported by a gas jet system from their production site through the biological shield to the experiment \cite{stend1980}. Besides the neutron-rich nuclides produced in the reactor, samples of transuranium elements up to \textsuperscript{252}Cf are also available.

The \textsc{Triga-Spec} setup consists of a Penning trap mass spectrometer, called \textsc{Triga-Trap}, and a collinear laser spectroscopy beamline, called \textsc{Triga-Laser}. They provide techniques to access model-independently nuclear ground-state properties, which are best suited to test nuclear theories. Nuclear (atomic) masses are determined via cyclotron frequency measurements of ions in the strong magnetic field of a Penning trap, while nuclear spins, magnetic and electric moments, and nuclear charge radii are extracted from laser spectroscopic investigations of hyperfine structure and isotope shifts.

For mass measurements, a double Penning trap mass spectrometer has been recently developed, which is situated in a similar superconducting magnet to \textsc{Shiptrap} \cite{raha2006}, \textsc{Jyfltrap} \cite{joki2006}, and \textsc{MllTrap} \cite{habs2006}. The main components of \textsc{Triga-Trap} are a cylindrical purification trap for beam preparation and a hyperbolical precision trap for the mass measurement \cite{webe2005,ferr2007}. Both are located in the bore of a 7-T superconducting magnet and are kept at cryogenic temperatures of 77\,K. The mass measurement can be performed with two different approaches: The destructive time-of-flight ion-cyclotron resonance (\textsc{Tof-Icr}) technique \cite{koni1995} is used for short-lived nuclides with half-lives $T_{1/2}<1\,$s, and the non-destructive image current detection (\textsc{Ft-Icr}) method \cite{mars1998} for rarely produced but rather long-lived species. The \textsc{Ft-Icr} method is a common technique used in chemistry in order to identify products of chemical reactions. For that, a minimum of 100-1\,000 ions of the identical species are required to obtain a \textsc{Ft-Icr} signal. We intend to apply this technique for the first time at a radioactive beam facility. For this purpose, the sensitivity has to be increased to a single-ion detection level. A resonator with a high quality factor $Q$ and the required signal processing electronics have been developed recently. In addition to single-ion detection in the precision trap, a non-destructive broad-band detection will be performed in the purification trap. This will allow to identify the trap content without the need to eject the stored ions. For the \textsc{Tof-Icr} measurements a double-detector setup will be used consisting of a conventional microchannelplate (MCP) and a channeltron with conversion dynode \cite{yazi2007}. The relative mass uncertainty $\delta m/m$ is expected to be in the order of $10^{-6}$ down to $10^{-7}$ for both measurement techniques.

Collinear laser spectroscopy \cite{kauf1976,anto1978} is a well-established tool for high-resolution optical spectroscopy. It is sensitive to nuclear ground-state properties, specially adapted for the investigation of short-lived radioactive nuclei, with a track record that started off from first experiments on rubidium and caesium isotopes performed at the \textsc{Triga} Mainz reactor \cite{schi1978,klem1979}. The concept of collinear laser spectroscopy is to overlap the fast beam of singly-charged ions at energies of typically 30\,-\,60\,keV with a continuous-wave (cw) laser beam in collinear or anti-collinear geometry. This technique allows studying short-lived ions and atoms, since they are investigated in-flight. Additionally, the spectroscopic resolution benefits from a compression of the velocity spread of the ion beam \cite{kauf1976,wing1976}. The use of a fixed-frequency laser is another advantage of this concept, since the frequency scanning is achieved by means of tuning the velocity of the ion beam and, thus, the resulting Doppler shifts.

In case the ionic spectrum is unfavorable for cw lasers, i.e.\,the transitions from the atomic ground-state are in the deep ultraviolet region, the ions can be neutralized in an alkali-vapour charge-exchange cell and, thus, spectroscopy can be performed on the neutral species. Differences of charge radii within an isotopic chain can be determined from measured isotope shifts (IS), whereas nuclear spins, magnetic dipole moments and electric quadrupole moments are extracted from the hyperfine structures (HFS). Several mechanisms can be used for the determination of an optical resonance \cite{geit2000}. Classical fluorescence detection by photomultipliers for isotope shift and hyperfine structure studies is a very general technique but moderately sensitive. Recently, its performance could be considerably improved using cooled and bunched ion beams \cite{niem2002}. Optical pumping with subsequent implantation and beta-asymmetry detection ($\beta$-NMR) is very sensitive and well suited for the determination of nuclear moments (see e.g.\,\cite{neye2005}). Furthermore, a variety of more specialized techniques were developed at the collinear laser spectroscopy setup \textsc{Collaps} at \textsc{Isolde} \cite{neug2000} including, e.g., collisional ionization processes with subsequent particle detection \cite{liev1992}.

The mass and laser spectroscopy branches at \textsc{Triga-Spec} will have two purposes. They will serve as a test station for the development of on-line mass and laser spectroscopic techniques (see below). In addition, the availability of some actinide isotopes and the coupling to the \textsc{Triga} Mainz enables measurements that provide interesting new data on masses or binding energies, charge radii, and hyperfine structures of nuclei in their ground and also isomeric states.  \textsc{Triga-Spec} will start with off-line measurements on actinide isotopes. Only few measurements on such isotopes were performed in the past, most of them with very low resolution. For actinium, protactinium, neptunium, and californium no direct mass measurements have ever been performed and no data on isotope shifts are available. Long-lived isotopes of these elements can be studied at \textsc{Triga} Mainz and are excellent test cases for the spectroscopy of even heavy and superheavy elements. Also hyperfine structure studies are of interest since many of the isotopes have not been investigated so far. In the case of on-line coupling, the fission products of $^{249}$Cf contain some very interesting cases for mass and laser spectroscopy. The region of the $N=50$ and $N=82$ shell closures can be studied, where the yield of $^{132}$Sn is about $1\cdot 10^{6}\,/\,$s, or the neutron-rich isotopes of silver, rhodium and ruthenium can be investigated. After a first round of on-line runs, more sensitive techniques like ion beam cooling and bunching as well as resonance ionization spectroscopy (RIS) will be applied to fully exploit the possibilities of isotope production at the reactor.

The benefit for future high-precision mass spectrometry and laser experiments from \textsc{Triga-Trap} and \textsc{Triga-Laser} will be primarily the development and test of devices and experimental procedures for the \textsc{Mats} Penning trap mass and
decay spectrometer and the LaSpec laser spectroscopy setup \cite{noer2006} at the low-energy branch of the super-fragment separator of \textsc{Fair}. On a shorter time scale, the \textsc{Shiptrap} facility \cite{raha2006} installed
behind the velocity filter \textsc{Ship} (\textsc{Gsi} Darmstadt) and dedicated to high-precision mass measurements on heavy and superheavy elements will benefit from the \textsc{Triga-Spec} setup since novel technical developments like the non-destructive single-ion \textsc{Ft-Icr} detection can be tested in detail first with heavy elements from uranium to
californium at \textsc{Triga} Mainz prior to the final installation at \textsc{Shiptrap}. The \textsc{Triga-Laser} setup will test new techniques to increase the accuracy of collinear laser spectroscopy. For instance, it will be checked if ion acceleration voltages up to 60\,kV can be determined with an accuracy better than $10^{-5}$. It is also planned to perform absolute frequency measurements of the investigated transitions by locking the laser system to a fiber-laser based frequency
comb \cite{kubi2005}, as an alternative to cyclic measurements between stable reference and radioactive isotopes. These techniques are mainly interesting for light and medium heavy isotopes, where accuracy is much more
important than for the heavier ones, and might later be more regularly employed for experiments like \textsc{Collaps} at \textsc{Isolde/Cern} or the collinear spectroscopy setup at \textsc{Isac/Triumf} in Vancouver.
Furthermore, it has recently been demonstrated that resonance ionization spectroscopy in a buffer-gas cell with
low-repetition high-power pulsed lasers can be used to obtain the very first information on the frequency of atomic transitions in a very heavy element such as fermium ($Z=100$) \cite{sewt2003,back2007}. Measurements with higher resolution, however, should probably be performed in ion traps since the production rates are too low to study them in flight.

\section{Basics of mass and laser spectroscopy for the determination of nuclear ground-state properties}

\label{sec_basics}

\subsection{Principles of Penning trap mass spectrometry}

Penning traps are the instruments of choice for direct mass measurements on stable as well as on short-lived nuclides with so far unsurpassed precision \cite{blau2006}. The high detection sensitivity enables investigations in rarely produced species at radioactive beam facilities. The principle of Penning traps is treated in more detail in Ref.\,\cite{brow1986}. By storing ions in a superposition of a strong homogeneous magnetic field, that defines the $z$-axis, and a weak electrostatic quadrupole field, their motion separates into three independent eigenmotions with the characteristic frequencies 
\begin{eqnarray}
\label{eigenfreq+}
\nu_+&=&\frac{1}{2}\left(\nu_c+\sqrt{\nu_c^2-2\nu_z^2}\right), \\
\label{eigenfreq-}
\nu_-&=&\frac{1}{2}\left(\nu_c-\sqrt{\nu_c^2-2\nu_z^2}\right), \\
\label{eigenfreqz}
\nu_z&=&\frac{1}{2\pi}\sqrt{\frac{qV_0}{mD^2}}.
\end{eqnarray}
Here, $\nu_+$ is the reduced cyclotron, $\nu_-$ the magnetron, and $\nu_z$ the axial frequency of an ion with a charge-to-mass ratio $q/m$ confined in a magnetic field $B$ and a trapping potential $V_0$. The parameter
\begin{eqnarray}
\label{D}
D&=&\sqrt{\frac{1}{2}\left(z_0^2+\frac{\rho_0^2}{2}\right)}
\end{eqnarray}
is defined for the ideal hyperbolical Penning trap by the minimum distances of the end caps $2z_0$ at $\rho=0$ and that of the ring electrode at $z=0$, i.e.\,$2\rho_0$. For a cylindrical trap an effective parameter $D^*$ can be
defined similarly. The eigenfrequencies of the ion motion are linked to the cyclotron frequency 
\begin{eqnarray} 
\label{nu_c}
\nu_c&=&\frac{1}{2\pi}\frac{q}{m}B
\end{eqnarray}
via the so-called invariance theorem \cite{brow1986} 
\begin{eqnarray} 
\label{invariance}
\nu_c^2&=&\nu_+^2+\nu_z^2+\nu_-^2,
\end{eqnarray}
or in the ideal Penning trap by 
\begin{eqnarray}
\label{sum_frequency}
\nu_c&=&\nu_++\nu_- .
\end{eqnarray}
A determination of the cyclotron frequency $\nu_c$ of the ion of interest and that of a reference ion ($\nu_{c,ref}$) leads to the atomic mass of the nuclide under investigation ($m_{atom}$) \cite{kell2003}. In case of singly charged ions the
relation is given by
\begin{eqnarray}
m_{atom}=\frac{\nu_{c,ref}}{\nu_c}\left(m_{ref}-m_e\right)+m_e.
\end{eqnarray}
Here, $m_{ref}$ is the mass of the reference ion used for calibrating the magnetic field and $m_e$ is the mass of the electron.

The frequency determination of a short-lived nuclide in a Penning trap is commonly carried out via the \textsc{Tof-Icr} method \cite{grae1980}: An electric quadrupolar rf field at the sum frequency $\nu_++\nu_-$ (cf. Eq.\,(\ref{sum_frequency})) is used to couple the two radial motions, leading to a periodic energy transfer between the two harmonic oscillators. Since $\nu_+\gg\nu_-$, the radial energy
\begin{eqnarray}
E_{rad}\propto\nu_+^2\rho_+^2-\nu_-^2\rho_-^2\approx\nu_+^2\rho_+^2
\end{eqnarray}
is dominated by the modified cyclotron motion. The energy $E_{rad}$ determines an orbital magnetic moment of the stored ions, which leads to a force
\begin{eqnarray}
\vec{F}=-\frac{E_{rad}}{B}\frac{\partial B}{\partial z}\vec{e_z}
\end{eqnarray}
on the ion in the magnetic field gradient when ejected from the trap. Thereby, the time of flight to a detector has a minimum in case the frequency used to excite the ions in the trap equals $\nu_c$.

Another possibility for the frequency determination is the \textsc{Ft-Icr} technique based on the image currents \cite{schw1991}
\begin{eqnarray}
i(t)=\frac{2\pi\nu_{ion}r_{ion}(t)q}{d}
\end{eqnarray}
the ions induce in the electrodes of the trap. In this simplified equation, $\nu_{ion}$ denotes one of the eigenfrequencies given in Eqs. (\ref{eigenfreq+},\,\ref{eigenfreq-},\,\ref{eigenfreqz}), $q$ is the charge-state, and $d$ is the modified electrode distance. A Fourier transformation of the time-domain signal reveals the ion frequency.

Most important for the applications mentioned above is the fact that the nuclear mass gives access to the nuclear binding energies via
\begin{eqnarray}
M(Z,N)=Z\cdot m_p+N\cdot m_n-\frac{E_B(Z,N)}{c^2},
\end{eqnarray}
where $(Z,N)$ denote the proton and neutron numbers, $(m_p,m_n)$ are the proton and neutron masses, and $E_B(Z,N)$ is the nuclear binding energy. Isomeric states with excitation energies in the order of a few ten to a hundred keV can be also resolved \cite{blau2004,webe2005a}.

\subsection{Nuclear ground-state properties from laser spectrosopy}

Optical techniques, applied to HFS or IS measurements, yield model-independent information on the nuclear structure \cite{otte1989,klug2003}. The theory of HFS and IS is sufficiently well understood to yield precise information on the size and shape of nuclei. On-line laser spectroscopy allows one to study the nuclear properties of ground-states of short-lived exotic isotopes which are available in only small quantities. The properties that can be studied are the nuclear spin $I$, the magnetic moment $\mu _{I}$, the spectroscopic nuclear quadrupole moment $Q_{s}$, and the change in the mean-square nuclear charge radii $\delta\langle r^{2}\rangle$ between isotopes. Experimental data can be determined with high accuracy and the nuclear parameters can be extracted without using a nuclear model. Since the model-independent data are collected in long isotopic chains reaching far from the valley of nuclear stability, these data provide clear information on single-particle as well as collective nuclear effects and enable stringent tests of nuclear models.

A detailed description about the extraction of nuclear properties from optical spectra is, for example, being given in \cite{otte1989} and will only be briefly summarized here: The transition frequency of an electronic transition shows a small shift between different isotopes of the same element. This is related to two terms: the difference in the nuclear mass leads to a difference in the nuclear center-of-mass motion, the so-called mass-shift (MS)
\begin{eqnarray}
\label{eq:mass_shift}
\delta \nu _{\mathrm{MS}}^{MM^{\prime}}=K_{\mathrm{MS}}\frac{M^{\prime}-M}{M^{\prime}M},
\end{eqnarray}
and the difference in the charge distribution $\delta\langle r^{2}\rangle$ gives rise to the volume or field shift (FS)
\begin{eqnarray}
\label{eq:field_shift}
\delta \nu _{\mathrm{FS}}^{MM^{\prime}}=\frac{2\pi Z}{3}\Delta\left\vert\Psi (0)\right\vert ^{2}~\delta \langle r^{2}\rangle
\end{eqnarray}
between the isotopes with masses $M$ and $M^{\prime}$. $K_{MS}$ is the mass shift constant and $\Delta\left\vert\Psi (0)\right\vert ^{2}$ is the change of the electron density at the nucleus in the eletronic transition. The mass shift has two contributions: the so-called normal mass shift, which is given by the reduced mass and easy to calculate, and the specific mass shift. The latter depends on the correlation between the electrons. Presently, this shift can only be calculated reliably for atoms or ions with at most three electrons. For the lightest elements, the nuclear volume is small but the relative difference in mass between the isotopes is large. Hence, the volume shift is only a tiny fraction of the mass shift, and the isotope shift has to be determined with high accuracy. In addition, highly accurate atomic structure calculations \cite{yan2000,drak2005} are required. This has so far only be accomplished for hydrogen \cite{hube1998}, helium \cite{muel2007}, and lithium \cite{sanc2006}. Contrary, for the heaviest isotopes it is easy to extract nuclear volume information since the field shift is the dominant part of the isotope shift and the mass shift can almost be neglected. Medium-resolution measurements are usually sufficient in these cases, but high efficiency is required to reach regions far away from stability. In the medium mass region there is more need for measurements with relatively high resolution, since mass shift and field shift cancel each other to a large degree.

The electromagnetic moments of isotopes with nuclear spin gives rise to the hyperfine structure splitting in optical transitions and its pattern, i.e.\,the number of lines and their relative strengths and positions are determined by the atomic and nuclear spins, $J$ and $I$, respectively. The spins, the magnetic dipole moment $\mu _{I}$, and the spectroscopic quadrupole moment $Q_{s}$ are directly connected to the hyperfine structure splitting by the Casimir formula 
\begin{eqnarray}
\label{eq:hfs}
\Delta \nu _{\mathrm{HFS}}=\frac{A}{2}C+\frac{B}{4}\frac{\frac{3}{2}C\left( C+1\right) -2I\left( I+1\right) J\left( J+1\right)}{\left( 2I-1\right)\left( 2J-1\right) IJ},
\end{eqnarray}
where $C=F\left( F+1\right) -I\left( I+1\right) -J\left( J+1\right)$, and $A,B$ are the magnetic dipole and the electric quadrupole coupling constant, respectively, given by
\begin{eqnarray}
A=\frac{\mu _{I}B_{e}(0)}{I\cdot J},\nonumber\\
B=eQ_{s}\frac{\partial ^{2}V}{\partial z^{2}}=eQ_{s}V_{zz}
\end{eqnarray}
with the magnetic field created by the electrons $B_{e}(0)$, and the electric field gradient $V_{zz}$ at the nuclear site.

The wealth of data on isotope shifts and hyperfine structures of short-lived isotopes has been obtained mainly by two laser techniques: Resonance Ionization Spectroscopy (RIS) \cite{alkh1983,sauv2000} of neutral atoms and Collinear Laser Spectroscopy (CLS) \cite{bill1995,neug2002} of neutral atoms or singly charged ions. In general, the first technique offers the highest sensitivity while the second provides higher resolution. The present limit for the minimum yield for collinear laser spectroscopy using bunched beams \cite{niem2002} or more specific high-sensitivity detection schemes \cite{liev1992} is of the order of 100\,ions/s, and the shortest-lived isotope investigated is \textsuperscript{11}Li with a nuclear half-life of $T_{1/2}=8$\,ms \cite{arno1987}. By resonance ionization spectroscopy, the HFS and the IS of radioactive atoms with a nuclear half-life as short as $T_{1/2}=1$\,ms (\textsuperscript{244f}Am) has been determined at a production rate of about 10\,ions/s \cite{back1998}. In principle, isotopes with shorter half-lives would still be accessible by laser spectroscopy, which depends only on the production yield and not on the nuclear half-life. Recently, also a magneto-optical trap was applied to determine nuclear charge radii of short-lived helium isotopes \cite{muel2007}.

\section{Experimental setup}

\label{sec_experimentalsetup} 
\begin{figure*}[tbp]
\begin{center}
\includegraphics* [width=0.8\textwidth]{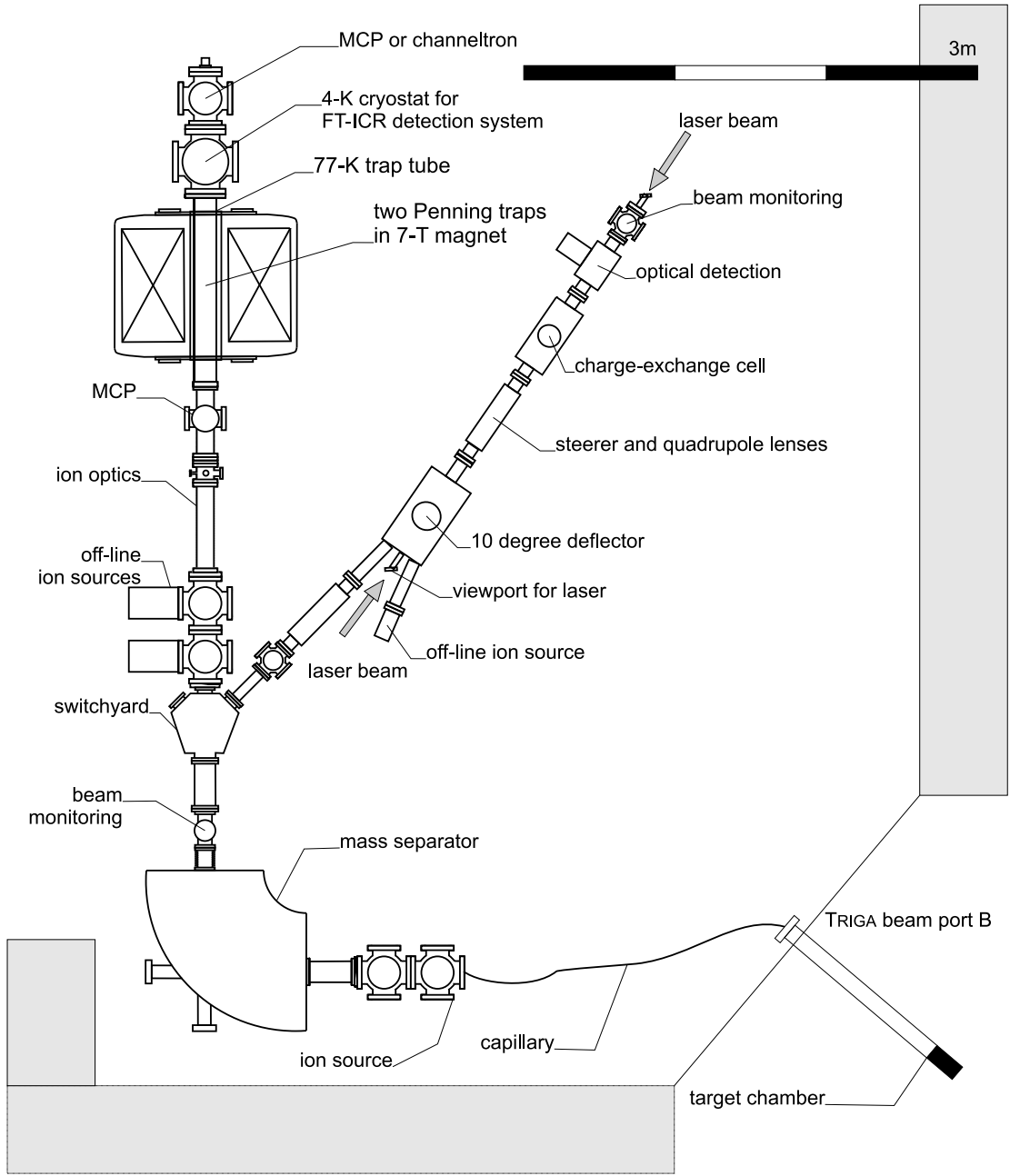}
\end{center}
\caption{The \textsc{Triga-Spec} setup with the mass spectrometry beamline to the left and the laser spectroscopy beamline to the right. For details see text.}
\label{fig_floorplan}
\end{figure*}
The experimental setup of \textsc{Triga-Spec} is presented schematically in Fig.\,\ref{fig_floorplan}. By fission of a heavy-element target, neutron-rich nuclides are created and transported to an on-line ion source via a gas jet system. A magnetic mass separator with a resolving power of about 1\,000 guides the ions of interest to a switchyard, where they are distributed to either the mass spectrometer \textsc{Triga-Trap} or the laser-spectroscopy setup \textsc{Triga-Laser}. In a next step, it is planned to install a radio-frequency quadrupole (RFQ) behind the mass separator at \textsc{Triga-Spec} to accumulate and bunch the ion beam coming from the on-line source \cite{herf2001,niem2001}. In addition, the emittance of the beam should be improved. For mass measurements, a double Penning trap system has been developed with traps operated at 77\,K. Two off-line ion sources provide the reference ions as well as the ions for experiments on elements heavier than uranium up to californium. For laser spectroscopy, the ion beams from the switchyard are bent by an electrostatic 10$^{\circ}$ deflector to spatially overlap them with the collinear laser beam. A subsequent charge-exchange and post-acceleration cell is used to tune the velocity of the ions in order to get them into resonance with the laser light. Optionally, the radioactive ions can be neutralized by passing them through alkali vapor inside the charge-exchange cell. For \textsc{Triga-Laser} an optical detection system employing photomultipliers is foreseen at the beginning. The major parts of both setups are described in more detail in the following sections.

\subsection{Isotope production, ionization and mass separation}

\subsubsection{The research reactor \textsc{Triga} Mainz}

The research reactor \textsc{Triga} Mainz is the only pulsable research reactor in Germany \cite{eber2000,hamp2006}. The presently installed 76 fuel elements are composed of an uranium-zirconium-hydrogen alloy with about 20\,\% \textsuperscript{235}U enrichment, where hydrogen serves as a moderator. Each cylindrical element has a length of 72.2\,cm and a diameter of 3.5\,cm. They are placed in a 20\,m$^{3}$ light water tank made of aluminum, which is surrounded by a concrete biological shield with four horizontal beam tubes. Close to the core a rotary specimen rack is placed for the simultaneous irradiation of up to 80 samples. In the steady-state mode, the reactor is operated at a thermal power of 100~kW, whereas, in the pulsed mode neutron pulses of 30~ms duration and a maximum peak-power of 250\,MW can be generated. This feature is due to the self-regulating fuel-moderator elements. When the temperature gets too high during a pulse, the moderation capability decreases leading to less fission events per time unit. Thereby, the reactor regulates itself back to normal operation within a fraction of a second. The pulsed mode enhances the production of very short-lived nuclides for physical and chemical experiments.

\subsubsection{Isotope production}

\begin{figure}[tbp]
\begin{center}
\includegraphics* [width=0.24\textwidth]{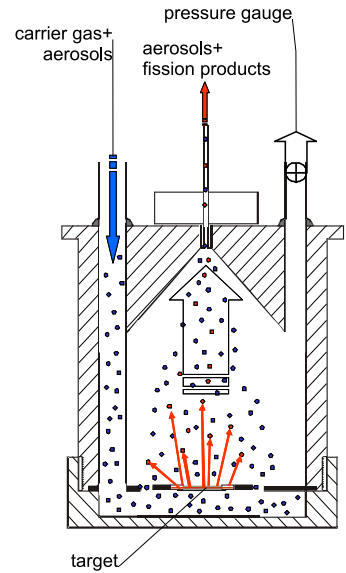}
\end{center}
\caption{Cross-section view of the target chamber for \textsc{Triga-Spec}. A thin target is fixed on one side of an aluminum or titanium chamber. Fission products recoil into the gas chamber, thermalize in the carrier gas and attach to aerosols. The gas stream transports the aerosols and fission products through a capillary to the ion source.}
\label{fig_targetchamber}
\end{figure}

For the \textsc{Triga-Spec} facility a target chamber (see Fig.\,\ref{fig_targetchamber}) will be placed close to the core of the reactor in beam port B. The fission target either consists of 300\,$\mu $g of \textsuperscript{249}Cf, 500\,$\mu$g of \textsuperscript{235}U or 500\,$\mu$g of \textsuperscript{239}Pu and is exposed to a thermal neutron flux of about $1.8\cdot 10^{11}$\,n/(cm$^{2}$s). The fission products recoil out of the target and are thermalized in a carrier gas, where either helium, argon, or nitrogen can be used. Before entering the target chamber, the carrier gas is loaded with aerosols with a particle size of 0.1\,-\,1\,$\mu $m. Suitable aerosol materials are inorganic salts like sodium chloride, potassium chloride, or lead dichloride. Alternatively, graphite aerosols can be used. The thermalized fission products in the target chamber are sorbed on the aerosols in the carrier gas and further transported through a polyethylene capillary with an inner diameter of 1\,-\,2\,mm. This gas jet transport system can provide transport times from the target to the experiment of less than 1\,s \cite{mazu1980}. The transport efficiency has been experimentally determined to be 60\,-\,70\,\% with different carrier gases and aerosol materials \cite{stend1980}. Thus, this transport system is suitable to get short-lived neutron-rich fission products to the on-line ion source of the \textsc{Triga-Spec} setup. 

In addition to the steady-state operation of the reactor, the pulsed mode can be used to enhance the production of the short-lived nuclides and thus suppress a background of the longer-lived species. With a Gaussian pulse of 250\,MW thermal peak power and a \textsc{Fwhm} of 30\,ms, the ratio between the activity produced in the pulse and the saturated activity produced in continuous mode for a nuclide of interest with half-life $T_{1/2}$ expressed in seconds is given by 55\,s\,/\,$T_{1/2}$ \cite{menk1975}. The pulsed mode can be used for the very short-lived nuclides as far as the reduced statistics due to the repetition rate is acceptable.

\subsubsection{The on-line ion source}

\begin{figure}[tbp]
\begin{center}
\includegraphics* [width=0.48\textwidth]{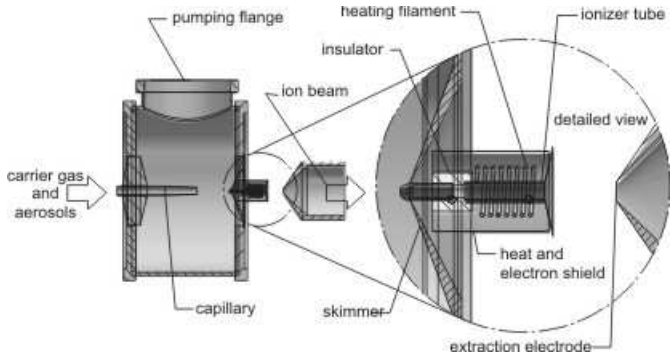}
\end{center}
\caption{Surface ion source as used for the \textsc{Helios} experiment at \textsc{Triga} Mainz \cite{mazu1980}. The part consisting of skimmer, ion source, and extraction electrode is enlarged on the right side. The aerosols with the attached fission products enter from the left. Through differential pumping and a skimmer most of the light carrier gas is separated from the aerosols and the attached fission products. The latter are surface ionized in the ionizer tube made of tungsten, which reaches temperatures up to 2300\,K by indirect heating. The ions are extracted and accelerated by a subsequent extraction electrode.}
\label{fig_onlinesource}
\end{figure}

The ion source system will have to separate the aerosol clusters loaded with fission products from the carrier gas, break up the clusters to release the nuclides of interest and ionize them efficiently. A sketch of an ion source that has been used previously for this purpose at the \textsc{Helios} separator \cite{brue1985} is shown in Fig.\thinspace \ref{fig_onlinesource}. In the first stage, the gas jet passes a differentially pumped skimmer system, where up to 90\,\% of the carrier gas are separated and pumped away by a roots pump. The cathode is heated by either electron bombardment or direct heating. In order to get a beam with a well-defined energy, the whole source region is held at a potential of 30-60~kV and the ions are extracted against ground with an extraction electrode. As the stability of the acceleration voltage is crucial especially for collinear
laser spectroscopy purposes, a high-voltage power supply with a stability better than 0.001\,\%\,$U_{\mathrm{nom}}$ in 8\,h will be used to offset the on-line ion source platform. Ionization efficiencies of around 6\,\% with the surface ion source described here and 0.5\,\% with a plasma ion source have been reached \cite{brue1985,brue1979}. The operation as a plasma source can be realized with a molybdenum ionizer tube and by generating an arc discharge inside the source initiated by applying a voltage difference to the separated metal parts of the source. Alternatively, the coupling of an ECR ion source is foreseen. 

\subsubsection{Mass separation and switchyard}

\begin{figure}[tbp]
\begin{center}
\includegraphics* [width=0.48\textwidth]{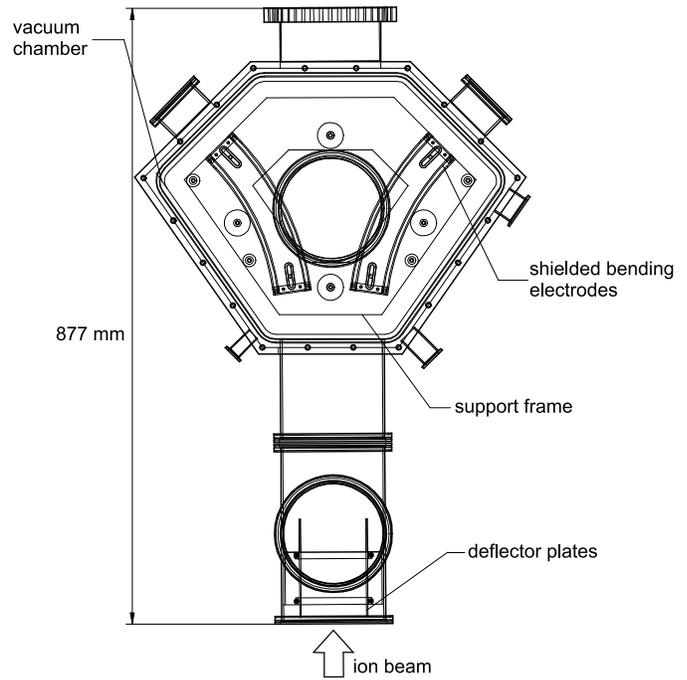}
\end{center}
\caption{Switchyard assembly. The ion beam can be transmitted straight or deflected by 45$^{\circ}$ to the right or the left. This is achieved by a combination of parallel deflector plates, which bend the ion beam by 8$^{\circ}$ and fringe-field shielded curved deflector plates.}
\label{fig_switchyard}
\end{figure}

A 90$^{\circ}$ dipole mass separator will be used for mass separation of the fission products. A mass energy product of 15\,MeV$\cdot$amu can be reached with a bending radius of 0.5~m and a magnetic field of up to 1.1\,T. This allows for acceleration voltages of 60\,kV for isotopes up to californium (250\,amu). For directing the mass separated ion beams to \textsc{Triga-Trap} or \textsc{Triga-Laser} an electrostatic 45$^{\circ}$ switchyard, similar to the one at \textsc{Isolde} \cite{kugl2000}, will be installed. As shown in Fig.\,\ref{fig_switchyard} it is a combination of an ion kicker, composed of two parallel plates, with two fixed sets of bending plates, which are shielded for fringe fields. The bending electrodes are permanently held at their deflecting voltage. The ions pass the switchyard straight or can be deflected by 8$^{\circ}$ by applying a potential to the parallel deflector plates so that the ions enter the curved bending electrodes. Fast switching between up to three experimental setups is possible without mechanical adjustment of the electrodes. The bending radius of the electrodes is 385\,mm and the gap 40\,mm. 

\subsection{The trap branch}

\subsubsection{The off-line ion sources}

\begin{figure}[tbp]
\begin{center}
\includegraphics* [width=0.48\textwidth]{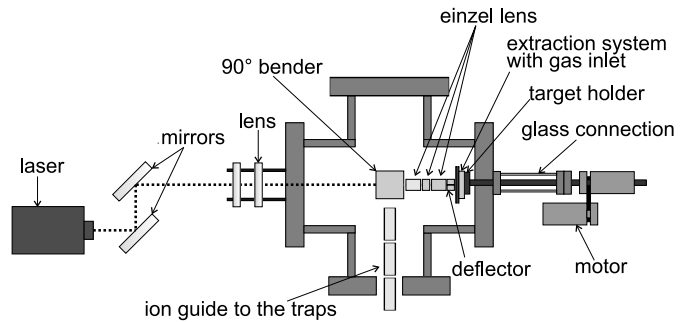}
\end{center}
\caption{Laser ion source for off-line experiments and calibration at \textsc{Triga-Trap}. A frequency-doubled Nd:YAG laser ($\protect\lambda=532$\,nm) is focused on the target, which consists of carbon material or the heavy element sample to be investigated. The ions are bent $90^\circ$ into the beam line.}
\label{fig_lasersource}
\end{figure}
\begin{sidewaysfigure*}
\begin{center}
\includegraphics*[width=0.98\textwidth]{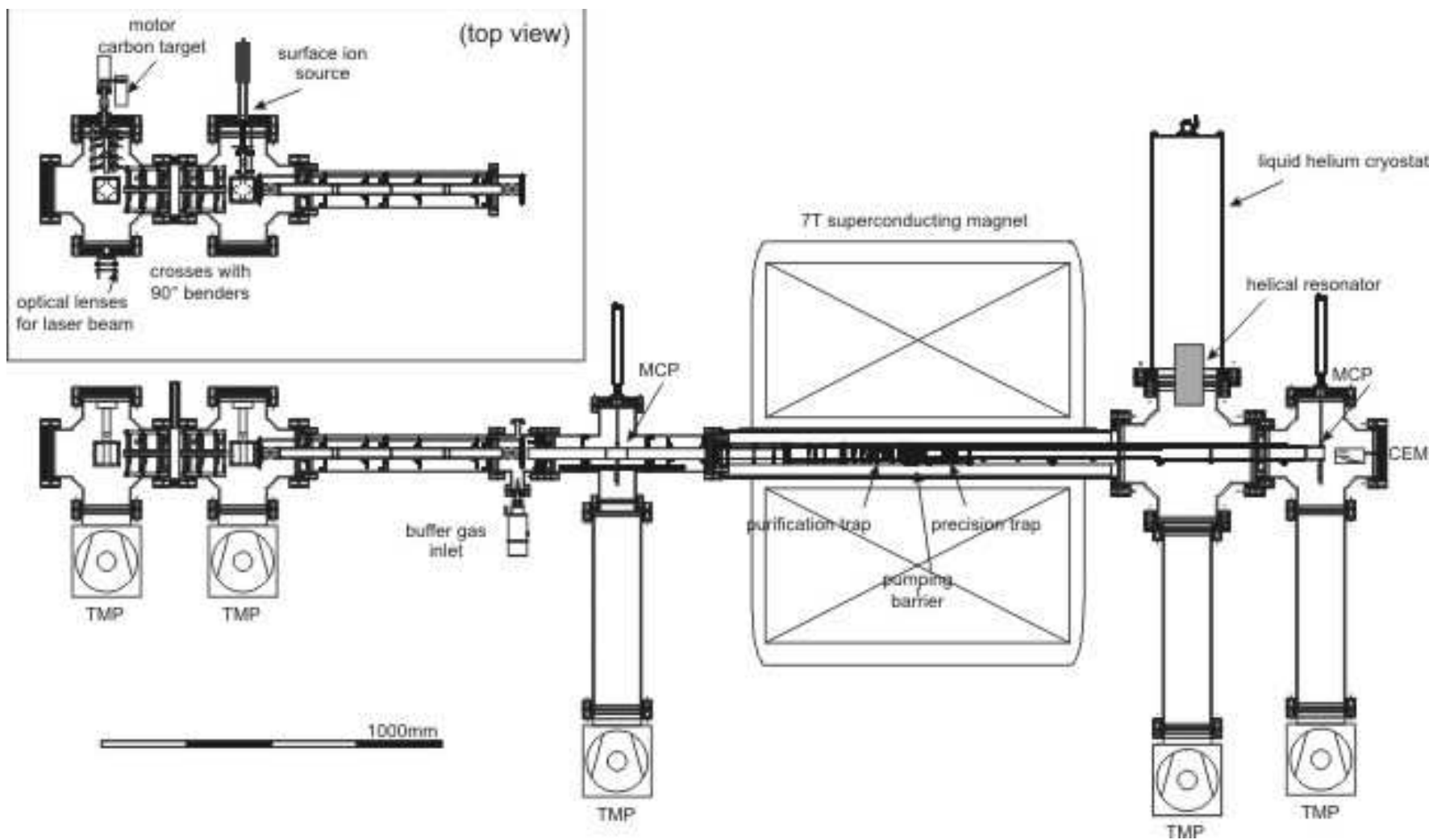}
\end{center}
\caption{The \textsc{Triga-Trap} double Penning trap setup. The two off-line ion sources on the left provide either alkali ions, carbon cluster ions or ions of heavy elements. The ions are guided by a system of einzel lenses and deflectors into the trap tube, which is centered in a 7-T superconducting magnet from \textsc{Magnex Scientific Ltd.}. The cylindrical purification trap and the hyperbolical precision trap are cooled to 77\,K by liquid nitrogen. A liquid helium cryostat houses the narrow-band \textsc{Ft-Icr} detection electronics including the helical resonator. The detectors for time-of-flight measurements as well as position-sensitive detection are mounted at the end of the setup. (TMP: turbomolecular pump, MCP: microchannel plate, CEM: channeltron electron multiplier.)}
\label{fig_setup}
\end{sidewaysfigure*}
The \textsc{Triga-Trap} setup contains two off-line ion sources for test purposes or calibration issues as well as for the production of heavy ions. A newly designed non-resonant laser ion source serves as a simple and reliable source for the creation of carbon cluster or heavy ions from samples of long-lived radio-nuclides. A pulsed frequency-doubled Nd:YAG laser with a wavelength of 532\,nm and a pulse length of 3\,-\,5\,ns is used for laser desorption and ionization. The repetition rate can be chosen between 1 and 15\,Hz, and the maximum energy in one pulse is about 50\,mJ at $\lambda=532\,$nm. The adjustment of the laser spot on the target is done with suitable mirrors (see Fig.\,\ref{fig_lasersource}). A target holder is mounted off-axis in order to allow rotation of the target and in this way a variation of the laser spot position on the target. The possibility to use the same source for the production of the ions of interest as well as the reference ions for calibration purposes is provided by using a target where half of the surface is covered with carbon material and the other half with a sample of the heavy element. A \textsc{Sigradur}\textsuperscript{\textregistered} or a C$_{60}$ target can be used to produce carbon cluster ions, which provide the mass calibration of choice across the entire nuclear chart. \textsc{Sigradur}\textsuperscript{\textregistered}is a solid-state form of carbon with a fullerene-like microstructure. The laser desorbs carbon clusters from the target surface and ionizes them. Due to the energy deposited by the laser pulse, some of the fullerenes are fragmented. In such sources as already installed at other trap facilities \cite{blau2003,kell2002a,chau2007}, typically a few hundred cluster ions are produced per laser pulse. The performance of the laser ion source for absolute mass measurements is discussed in Sec.\,\ref{ssec_lasersource}.

The second off-line ion source is a surface ion source designed by R. Kirchner \cite{kirc1981}. The material to be ionized is embedded in a finely ground substance of ceolite ceramics. It is surface ionized in a tungsten tube heated by electron bombardment. This surface ion source provides alkali ion beams with a current of up to a few nanoamperes for mass calibration. 

\subsubsection{The ion optics}

Figure\,\ref{fig_setup} shows the detailed setup of the mass spectrometer. Simulation studies have been carried out for all ion optical elements with the ion trajectory calculation programm \textsc{Simion} 8.0 \cite{sim} indicating a transport efficiency of about 80\,\% through the whole setup. The ion transport optics between the off-line sources and the traps consists mainly of cylindrical electrodes grouped to einzel lenses and drift sections. Each ion source has an extraction plate with a 2\,mm hole to limit the emittance of the extracted ion beam. A short einzel lens provides a focus in the center of the following $90^{\circ}$ bender guiding the ions into the main beam line. This arrangement enables on-line experiments and suppresses the neutral atom background. After beam bending an aperture with a 3\,mm hole ensures that the ions are injected on axis into the following transport section. Furthermore, this aperture improves the beam quality in case of the off-line ion sources. A pair of deflectors allows for displacement and bending of the beam. The deflectors are made of a cylinder cut longitudinally into four equal segments to which appropriate voltages are applied. Two einzel lenses focus the beam prior to a second pair of deflectors, which can be used to readjust the beam position and angle again. A third einzel lens allows proper forming of the ion beam for the injection into the traps through the magnetic field gradient of the superconducting 7-T magnet (compare Fig.\,\ref{fig_onaxiszoomed}). A movable microchannelplate (MCP) detector is used at the center of this lens to optimize the ion optics parameters prior to the injection into the magnetic field.

\subsubsection{The superconducting magnet}

\begin{figure*}[tbp]
\begin{center}
\includegraphics* [width=0.7\textwidth]{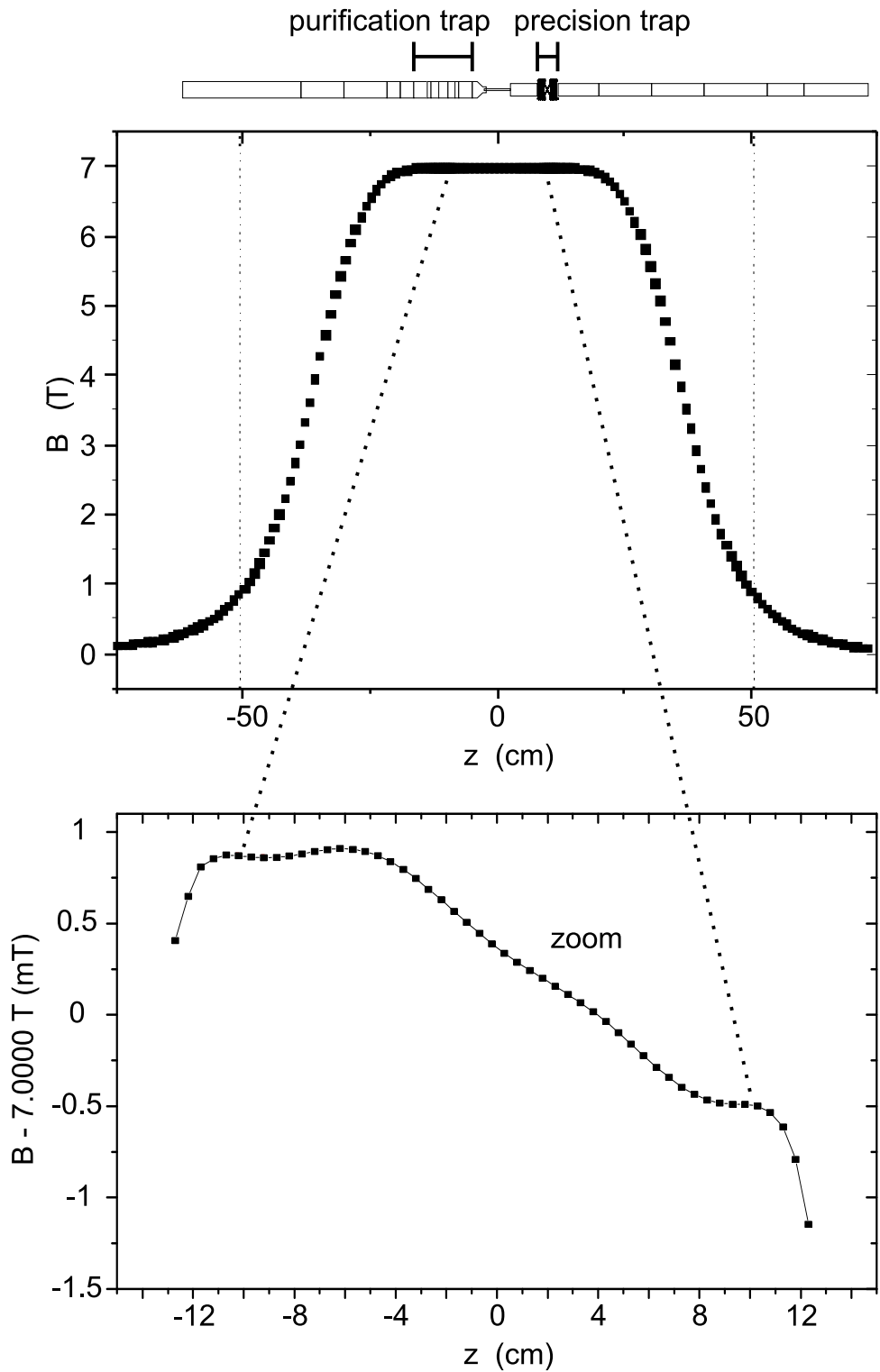}
\end{center}
\caption{Magnetic field of the superconducting magnet on axis of the bore. The electrode geometry including both Penning traps is shown on the top. The two homogeneous regions with $\Delta B/B =0.6$\,ppm for the purification trap and $<\,0.3$\,ppm for the precision trap at $\pm 10$\,cm, respectively, are zoomed in.}
\label{fig_onaxiszoomed}
\end{figure*}

Both Penning traps are placed in the bore of the same superconducting magnet manufactured by \textsc{Magnex Scientific Ltd.}. It is a 7-T magnet with a 160\,mm room-temperature bore similar to the ones at \textsc{Shiptrap} \cite{raha2006}, \textsc{Jyfl-Trap}, and \textsc{Mll-Trap} \cite{habs2006}. The magnet is actively shielded, which reduces the fringe field at the position of the collinear laser beamline and enables as well the operation of turbomolecular pumps close to the traps. The on-axis plot of the magnetic field strength for a distance of 24\,cm around the geometrical center of the magnet (Fig.\,\ref{fig_onaxiszoomed}) shows the two very homogeneous regions for the two Penning traps. The homogeneity $\Delta B/B$ on a circle with 5\,mm radius around the purfication trap center has been determined with a NMR probe to be about 0.6\,ppm. For the precision trap, the value is better than 0.3\,ppm. The decrease of the magnetic field due to the flux creep effect \cite{beas1969}, which is inherent to superconducting magnets, is corrected for by an additional coil resulting in a field drift $\Delta B/(B\cdot \Delta t)<10^{-10}\,/\,$h.

\subsubsection{The Penning traps}

Similar to many other on-line Penning trap setups for mass measurements \cite{blau2006}, a first trap (purification trap) is used for the ion cloud preparation, such as cooling and isobaric separation, and a second one (precision trap) for the actual mass measurement. The complete electrode stack including both traps as well as the injection and ejection drift tubes
is placed in a 77-K vacuum tube.
\begin{figure*}[tbp]
\begin{center}
\includegraphics* [width=0.8\textwidth]{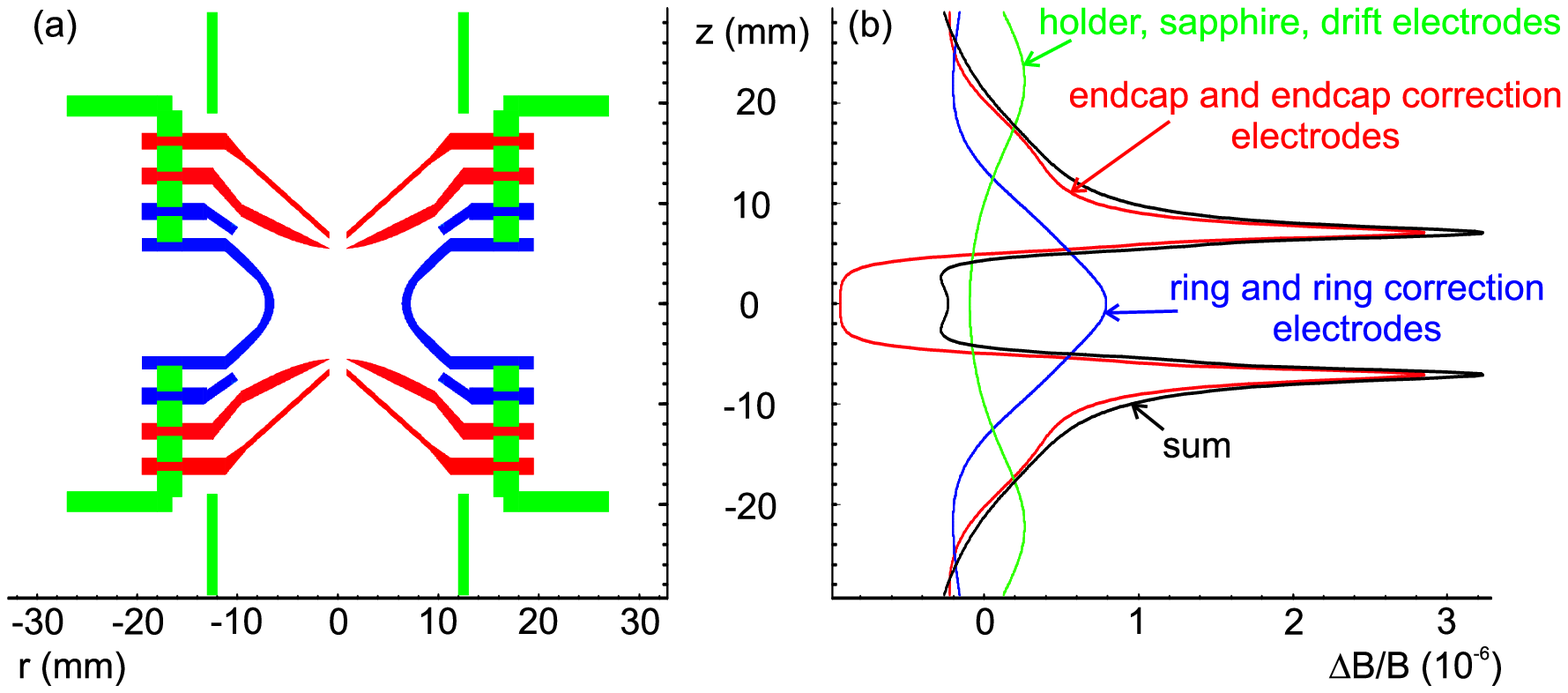}
\end{center}
\caption{(Color) (a) Hyperbolic precision trap with ring and inner correction electrodes (blue), endcaps and outer correction electrodes (red), isolated by sapphire rings (green). Holders and drift tubes are also indicated in green. (b) The relative change of the magnetic field due to the magnetization of the different materials used to construct the trap. The black line sums up all contributions.}
\label{fig_traps}
\end{figure*}
The purification trap is a cylindrical seven-electrode Penning trap with an inner diameter of 32\,mm and a total length of 212.5\,mm. The ion confinement is provided by the superposition of a strong homogeneous magnetic field defining the $z$-axis and a weak electrostatic quadrupole field. The latter is generated by a stack of cylindrical electrodes, parallel to the magnetic field lines. The basic properties of Penning traps are described in detail in \cite{brow1986}. The purification trap consists of seven electrodes: two endcaps, two endcap correction electrodes, two ring correction electrodes, and the ring electrode split into two $40^\circ$ and two $140^\circ$ segments. The smaller ones are dedicated to the excitation of the ion motion whereas the larger segments serve as pick-up electrodes for the
induced image current. In addition, the endcap correction electrodes are split into equal halves to enable axial excitation and detection as well. At the trap entrance, two more cylinders similar to the endcaps can be used to modify the axial potential. A funnel electrode is placed at the ejection side. In the purification trap, the ions coming from the source are cooled
by collisions with helium buffer-gas at a pressure of about $10^{-4}$\,mbar. By applying rf fields with the appropriate cyclotron frequencies, the ions under investigation can be mass-selectively centered in the trap \cite{sava1991}. When ejected, only the centered ions can pass the pumping barrier (Fig.\,\ref{fig_setup}) used for differential pumping. This is a channel with a diameter of 1.5\,-\,3\,mm placed between the two traps \cite{neid2007}. It ensures a pressure difference of $>100$ by reducing the helium flow from the first to the second trap. Whereas a high buffer-gas pressure is required to cool the ions in the purification trap, the vacuum conditions should be as good as possible in the precision trap. Ion collisions with residual gas atoms damp the ion motion, and thus lead to a reduced accuracy in the determination of the mass value \cite{kret2008}.

The precision trap (see Fig.\,\ref{fig_traps}\,(a)) is a hyperbolic Penning trap. The ions are injected and ejected through 1.8\,mm holes in the endcaps and the endcap correction electrodes. The trap is embedded into a system of cyclindrical drift tubes, which also serve as the time-of-flight section, when the destructive ion detection is used. The precision trap has seven electrodes and the segmentation of the electrodes is done similar to the purification trap, since a non-destructive image current detection will be performed here as well. The hyperbolic shape of the electrodes was chosen to minimize higher-order terms in the electric potential in a large volume. The inner diameter of the ring electrode is $2\,\rho_0=12.76\,$mm. The ratio between the radius of the ring electrode $\rho_0$ and the distance of the endcaps $2\,z_0$ is chosen as $\rho_0/(2\,z_0)\approx 1.16$. An orthogonalized trap can be created \cite{gabr1983}, in which the ion frequencies are practically independent of the motional amplitudes. The shape of the electrodes has been optimized to reduce the influence on the magnetic field and, thus, to minimize $\Delta B/B$ in the trap due to the magnetization of the materials used. For the electrodes, oxygen-free copper was chosen. Figure\,\ref{fig_traps}\,(b) shows the calculated contributions of each electrode as well as the resulting magnetic field inhomogeneity \cite{webe2004}.

The ion detection system is attached to the large segments of the ring electrode to determine the modified cyclotron frequency related to the mass. A small trap size was chosen to obtain the highest possible detection efficiency without a significant detoriation of the ion motion. Upon excitation of the radial motion to $0.8\,\rho_0$ in this hyperbolic trap, ions will still probe the very homogeneous region of the magnetic field, which directly determines the achievable uncertainty in the frequency determination.

\subsubsection{Detection of the cyclotron resonance}

\label{sec_iondetection} 
\begin{figure}[tbp]
\begin{center}
\includegraphics* [width=0.47\textwidth]{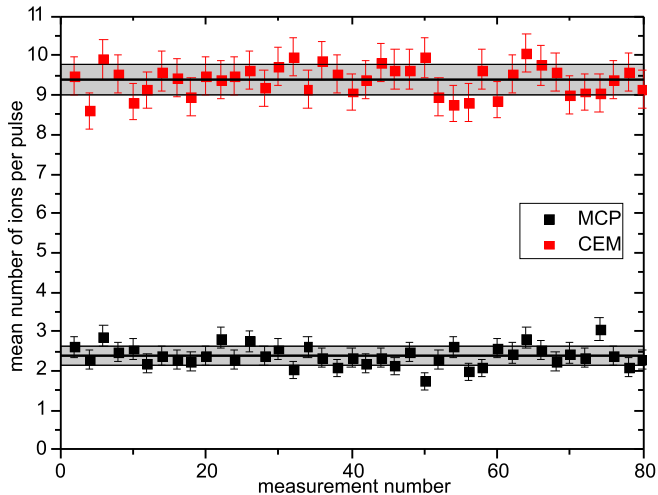}
\end{center}
\caption{Comparison of the detection efficiency of a microchannelplate (MCP) with that of a channeltron detector (CEM) for an identical number of incoming ions. The lines and the grey bands indicate the mean values and their uncertainties. A 3.5 times higher detection efficiency is observed for the CEM.}
\label{fig_mcpcem}
\end{figure}

The cyclotron resonance is observed in the destructive TOF mode by ejecting the ions from the precision trap and by recording the time of flight from the trap to the detector as a function of the applied radiofrequency excitation. In the non-destructive \textsc{Ft-Icr} mode, the image current induced by the ions in the ring electode is observed. Microchannelplates (MCP) \cite{wiza1979} and channeltrons (CEM) are commonly used for TOF detection. These destructive ion detectors are based on the secondary-electron emission and multiplication. In the case of MCP detectors, usually two plates are used in a stack. This Chevron configuration reaches an efficiency of about 30\,-\,50\,\% for ion energies around 2\,keV. At \textsc{Triga-Trap} a channeltron detector will be used in addition to the MCP \cite{yazi2007}. This detector can reach detection efficiencies $\geq$90\% in combination with a conversion dynode (Fig.\,\ref{fig_mcpcem}). A similar detector setup is in use at \textsc{Isoltrap} \cite{yazi2007} and \textsc{Shiptrap}.

\begin{figure}[tbp]
\begin{center}
\includegraphics* [width=0.47\textwidth]{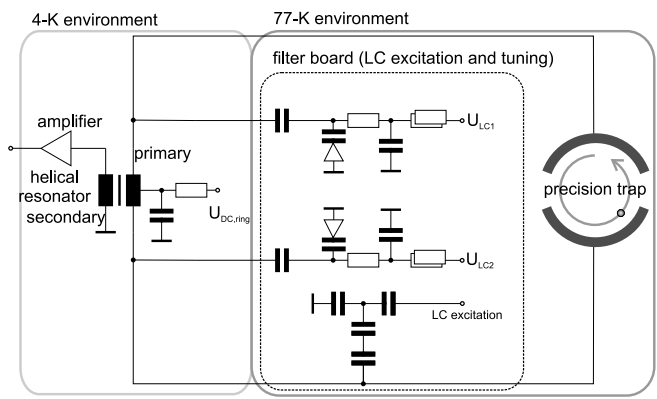}
\end{center}
\caption{The narrow-band \textsc{Ft-Icr} detection circuit. The ions induce an image current in the segments of the ring electrode of the precision trap. This current is guided to a helical resonator filter circuit, which creates a voltage drop corresponding to the current. The voltage is amplified in several stages prior to a discrete Fourier transformation revealing the frequencies of the ion motion in the trap. Some additional circuitry to tune the resonance frequency of the filter is shown as well.}
\label{fig_fticr}
\end{figure}

The detection circuit for the \textsc{Ft-Icr} mode is shown in Fig.\,\ref{fig_fticr}. For the first time, a single-ion detection by induced image currents in the trap electrodes will be used in an on-line Penning trap mass spectrometer for short-lived nuclides. This technique is well known in chemistry as the Fourier Transform-Ion Cyclotron Resonance (\textsc{Ft-Icr}) method, but is used there with typically several hundreds of simultaneously stored charged particles \cite{mars1998}. For \textsc{Triga-Trap}, an advanced system called narrow-band \textsc{Ft-Icr} has been developed, which can detect the induced image currents of a single ion. The signal is in the order of a few ten to a few hundred femto amperes. Non-destructive ion detection is also used at Penning trap mass spectrometers for stable isotopes, e.g.\,\textsc{Mit-Trap} (now: \textsc{Fsu}, Tallahassee, USA) \cite{rain2004,shi2005}.

At \textsc{Triga-Trap}, the \textsc{Ft-Icr} system is attached to the two larger segments of the ring electrode of the precision trap to detect the modified cyclotron frequency $\nu_+$ (see Eq.\,(\ref{eigenfreq+})). The mass is determined via the relation 
\begin{equation}
\frac{q}{m}=\frac{\nu_+^2}{a\nu_++b}\,,
\end{equation}
where $a=B/(2\pi)$ and $b=V_0/(8\pi^2D^2)$ are parameters related to the magnetic field $B$, the trapping potential $V_0$ and the trap geometry $D$ (see Eq.\,\ref{D}). The parameters $a$ and $b$ have to be calibrated by two reference measurements, e.g.\,with carbon clusters close to the mass of the ion of interest, prior and after the measurement with the ion of interest.

\begin{figure*}[tbp]
\begin{center}
\begin{tabular*}{\textwidth}{@{\extracolsep{\fill}}cccc}
(a) & \includegraphics* [width=0.47\textwidth]{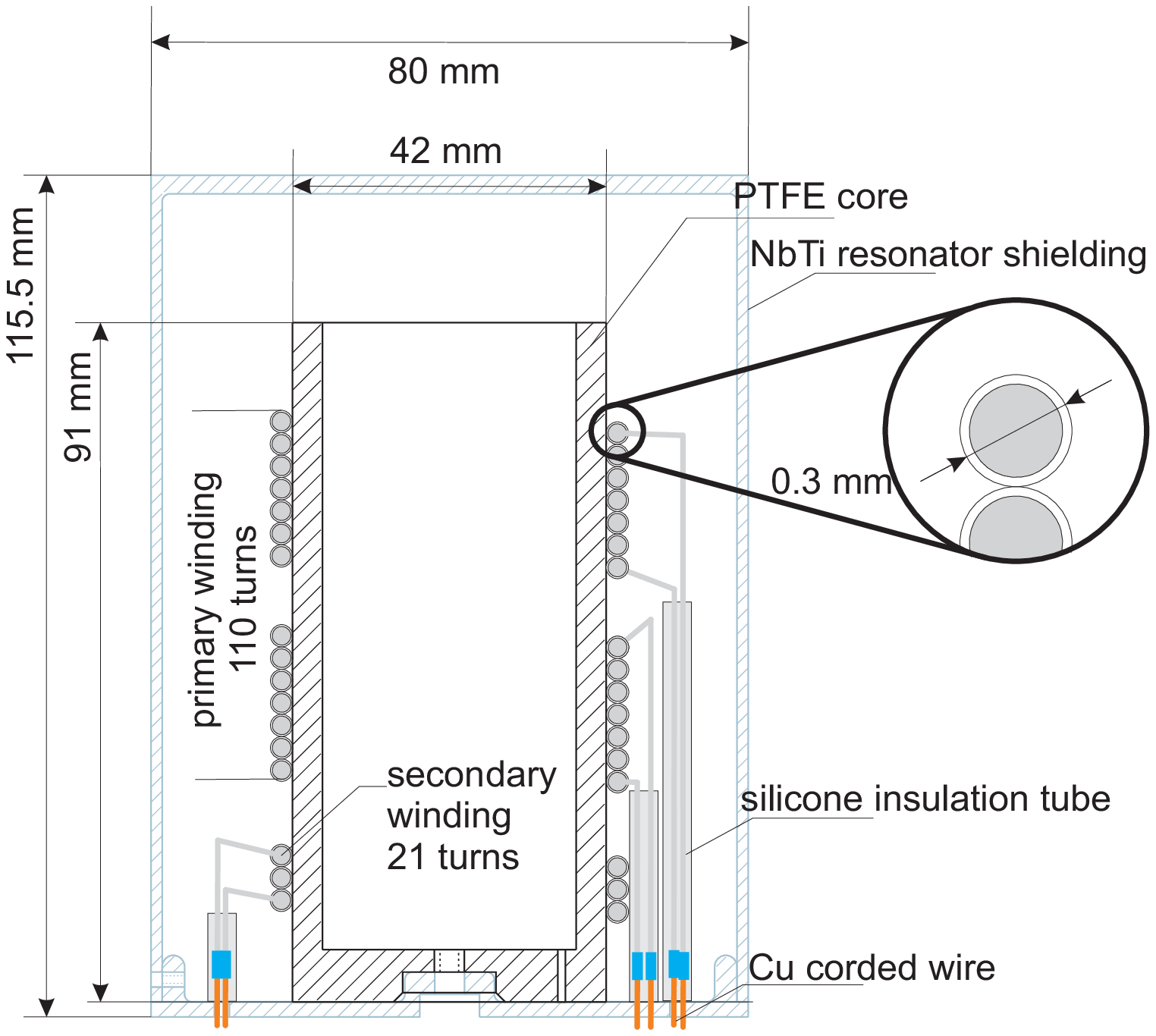} & (b) & %
\includegraphics* [width=0.4\textwidth]{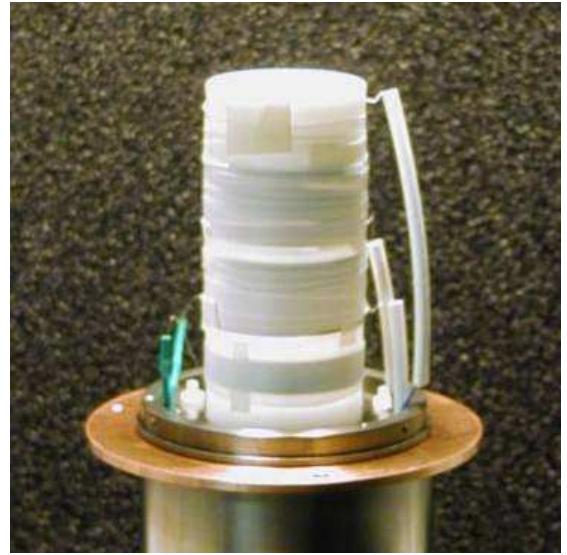}%
\end{tabular*}%
\end{center}
\caption{(a) The superconducting helical-coil transformer for narrow-band FT-ICR detection. It consists of three different coils with NbTi wire. The two upper coils form the primary side with a center tap, the lower package the secondary one of a transformer which is used to couple the subsequent amplifiers. The transformation ratio is 110:21. The wire is wound around a PTFE core and additionally fixed with PTFE tape to stabilize the coil at low temperatures. A NbTi shield completes the helical transformer in operating condition. (b) Photograph of the coil.}
\label{fig_resonator}
\end{figure*}
\begin{figure*}[tbp]
\begin{center}
\begin{tabular*}{\textwidth}{@{\extracolsep{\fill}}cccc}
(a) & \includegraphics* [width=0.4\textwidth]{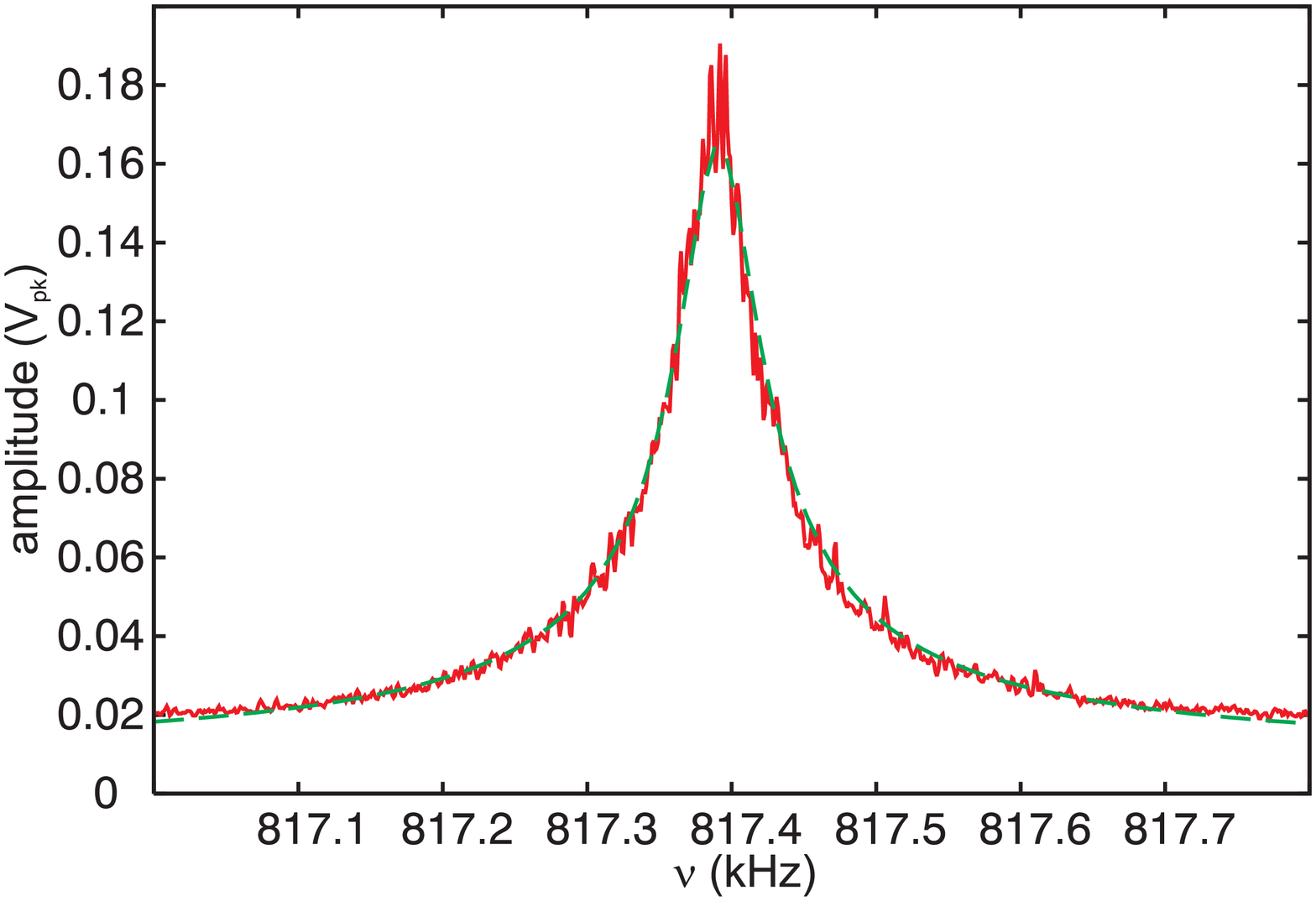} & (b) & %
\includegraphics* [width=0.4\textwidth]{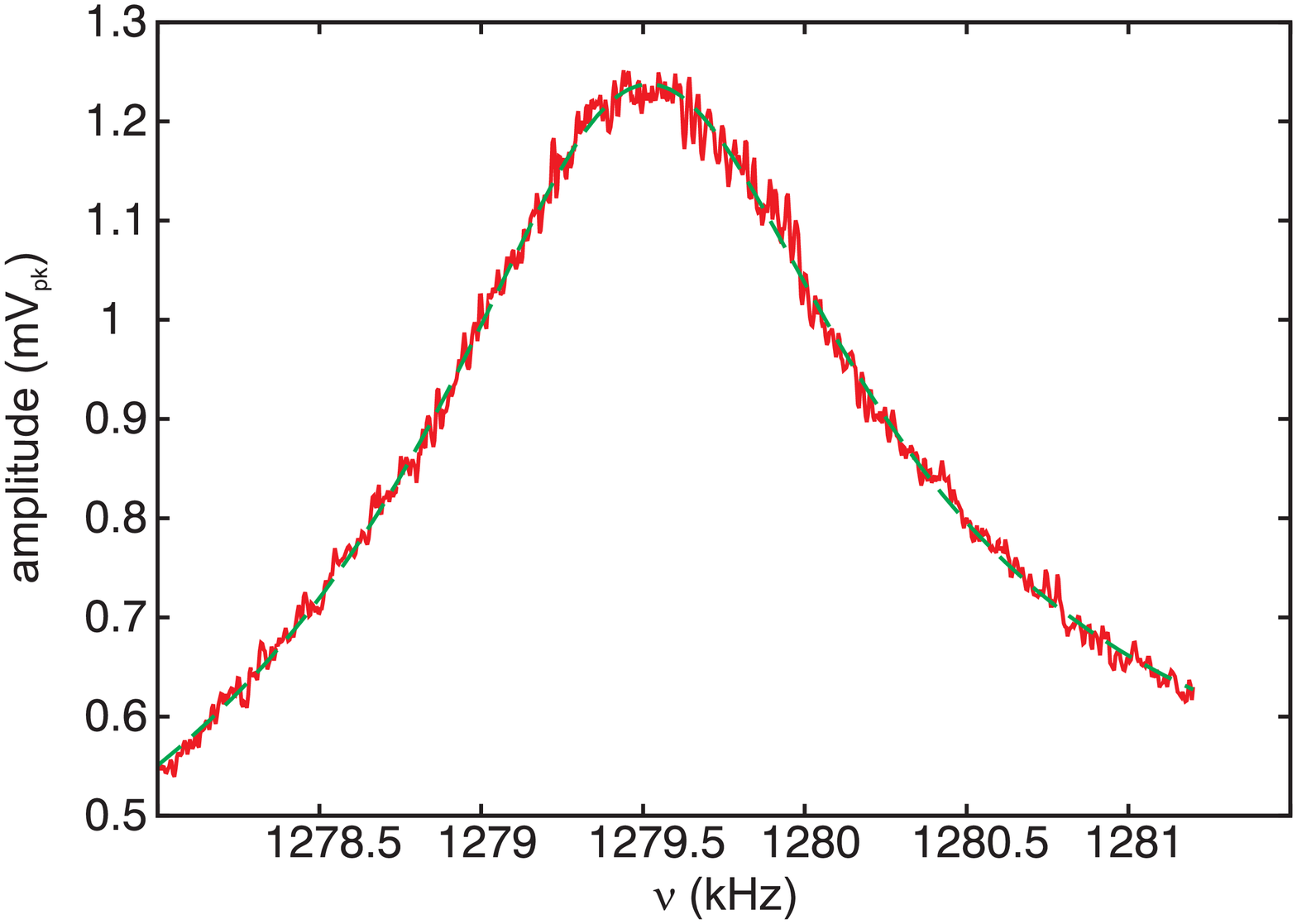}%
\end{tabular*}%
\end{center}
\caption{Resonances of the superconducting helical coil: (a)A capacitor of 106\,pF is connected to the coil and determines the central frequency in this case. A fit of the theoretical line shape (dashed line) reveales a $Q$-value of $15\,372\pm 153$. (b)The trap is connected to the resonator and the $Q$-value is $1\,005\pm 7$.}
\label{fig_qresonance}
\end{figure*}
The detection circuit attached to the pick-up electrodes mainly consists of a superconducting helical-coil transformer (see Fig.\,\ref{fig_resonator}) forming a parallel resonance circuit in combination with the parasitic capacitances of the trap and the wires. Thereby, the detection bandwidth is reduced to a few hundred Hz, which improves the signal-to-noise ratio. Further noise reduction is achieved by operating the trap in a 77-K environment. The helical coil consists of three packages of NbTi wire windings. Two of them form the primary side, the third one the secondary side of a transformer with a winding ratio of 110:21. The primary side has a center tap to float the coil with the DC voltage for the ring electrode segments which is required for the trapping potential. The wire is wound around a hollow \textsc{Ptfe} core and is fixed with \textsc{Ptfe} tape to stabilize the coil at low temperatures. The core is mounted in a cylindrical NbTi shield, which completes the helical resonator. Superconductivity of NbTi is obtained by connecting the resonator with a copper plate to a liquid-helium cryostat at 4.7\,K \cite{webe2004}. Off-line tests of the performance of the narrow-band \textsc{Ft-Icr} detection system resulted in an unloaded quality factor of the resonator of about $Q=$15\,000. The intrinsic capacitance of the pure resonator is $(6.3\pm 1.8)$\,pF, its inductance $(337\pm 6)\,\protect\mu$H. With the trap as well as all electronics connected, $Q$ is still in the order of 1\,000 (see Fig.\,\ref{fig_qresonance}). These results meet the requirements for single-ion sensitivity. The resonance frequency can be shifted with varactor diodes over about 10 atomic mass units in the present configuration with a resonance frequency matched to the stable rubidium isotopes. Within this mass range it is possible to keep the $Q$-value nearly constant at $Q\approx 1\,000$.

A transistor amplifier is directly connected to the secondary side of the transformer and kept also at 4.7\,K. It feeds an operational amplifier at room temperature and creates a differential output signal. The last stage before sampling the signal is a heterodyning \textsc{Ft-Icr} amplifier with a low-pass filter. Its main part is a mixer shifting the signal frequency (500\,kHz\,-\,1\,MHz) to frequencies below 10\,kHz. The total gain can be adjusted to the needs and is currently set to about $10^5$. The analog signal is sampled prior to discrete Fourier transformation. The analog-to-digital converter (ADC) has a limited number of samples in a transient, which gives the transient length in combination with the sampling rate. The frequency-domain bin width in a discrete Fourier transformation is given by the inverse of the transient
length. To increase the resolution, i.e. to reduce the frequency-domain bin width, the transient has to be prolonged, which can only be achieved by a reduction of the sampling rate due to limited memory. The Nyquist theorem states that the sampling rate has to be twice as large as the highest frequency component in the signal in order not to lose any information.

Besides the narrow-band \textsc{Ft-Icr} detection with single-ion sensitivity in the precision trap, \textsc{Triga-Trap} also features a broad-band image current detection in the purification trap to investigate the trap content without the need to eject the ions. The required number of ions for a signal is estimated to be in the order of 1\,000. Contaminations
can be cleaned away by a mass selective excitation of the ions' motion. In the broad-band method, the image current is directly fed into a low-noise amplifier for each of the two segments of the ring electrode. A differential amplification stage followed by the required heterodyning completes the analog signal processing. The sampling and the discrete Fourier transformation are done by one National Instruments data acquisition card \textsc{Ni Pci}-4551 for both, the broad-band and the narrow-band \textsc{Ft-Icr} detection. This card provides two analog inputs with a maximum input frequency of 102.4\,kHz. The discrete Fourier transformation is performed using a Fast Fourier Transformation (FFT) algorithm.

\subsubsection{Data acquisition and control system}

The \textsc{Triga-Spec} setup is controlled via the Lab\textsc{View} based CS control system developed at \textsc{Gsi} Darmstadt \cite{beck2004} which controls most of the electronic devices including the timing of the experimental procedure and
triggers the readout of the data acquisition as well as the archiving of the data. To balance the load of the controlling \textsc{Cpu}s, the system is distributed via an ethernet connection over several computers. Additionally,
the MM6 graphical user interface from \textsc{Lebit} \cite{ring2006} is used to set the parameters for the mass measurement cycle and to visualize the experimental data. Thereby, all typical tasks like beam transport optimization by voltage scans, frequency and amplitude scans, as well as time-of-flight mass measurements can be performed. The integration of the non-destructive image-current detection scheme into the data acquisition for \textsc{Triga-Trap} is planned.

\subsubsection{Accessible half-lives for mass measurements}

In the TOF mode, the main limiting factor concerning the accessible half-lives is the time for transportation of the short-lived nuclides from the target chamber in the reactor to the ion source. Here, transport times in the order of 1\,s are expected \cite{mazu1980}. The transport between the ion source and the Penning traps is well below 1\,ms and therefore negligible. Purification in the first Penning trap takes typically 100\,ms. The ions are then stored in the precision trap and excited for 100\,ms up to 1\,s for a TOF measurement, depending on the aimed resolving power and precision, as well as on the half-life of the nuclide of interest. In the \textsc{Ft-Icr} mode, the observation time of the induced image currents is the main limiting factor. This time will be typically in the order of 1\,-\,10\,s to get a sufficient transient length of the time-domain signal. In total, the time-of-flight resonance technique will allow mass measurements on short-lived nuclides with half-lives down to 200\,ms (depending on the yield and the decay losses), whereas the  \textsc{Ft-Icr} technique can be used for rather long-lived species starting at 5\,s and longer half-lives.

\subsubsection{Absolute mass measurements and heavy-ion production}

\label{ssec_lasersource}
\begin{figure}[tbp]
\begin{center}
\includegraphics* [width=0.47\textwidth]{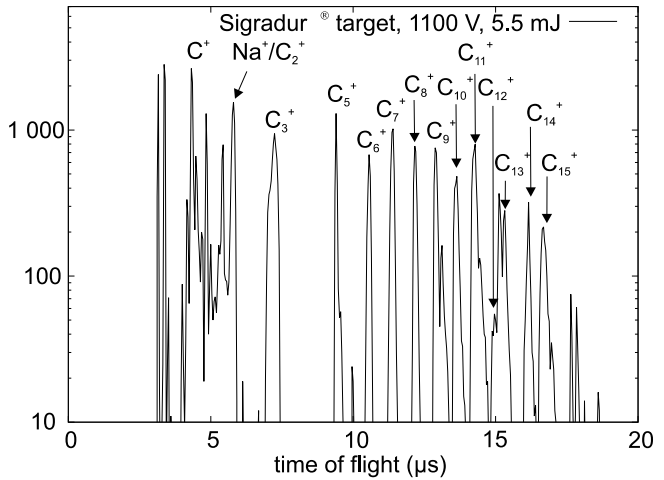}
\end{center}
\caption{Cluster spectrum produced with a \textsc{Sigradur} \textsuperscript{\textregistered} target and a frequency-doubled Nd:YAG laser. Carbon cluster ions up to C$_{15}^+$ are obtained.}
\label{fig_sigradure}
\end{figure}

\begin{figure}[tbp]
\begin{center}
\includegraphics* [width=0.47\textwidth]{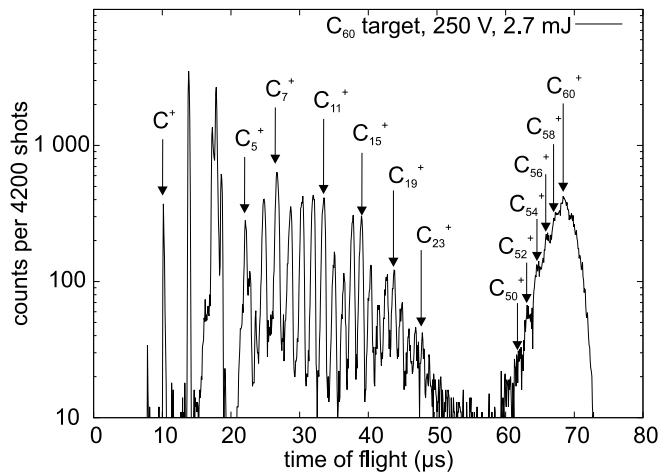}
\end{center}
\caption{Carbon cluster spectrum produced with a C$_{60}$ target and a frequency-doubled Nd:YAG laser. Cluster ions up to C$_{60}^+$ could be identified.}
\label{fig_c60}
\end{figure}

\textsc{Triga-Trap} will carry out absolute mass measurements by using carbon clusters as mass references \cite{blau2002,blau2003a}. Clusters up to C$_{23}^+$ were produced in pilot measurements with a \textsc{Maldi-Tof} system \cite{blau2003}. In a test setup, the performance of the laser ion source has been investigated in detail with a \textsc{Sigradur}\textsuperscript{\textregistered} target at 4\,-\,6\,mJ laser pulse energy (Fig.\,\ref{fig_sigradure}). Carbon cluster ions up to C$_{15}^+$ are produced and detected. Some impurities on the target and residual gas in the vacuum chamber are ionized as well and occur in the spectrum. During the tests with the \textsc{Triga-Trap} laser ion source the size of the laser focus on the target turned out to be a crucial parameter. In addition to the \textsc{Sigradur}\textsuperscript{\textregistered} target the use of C$_{60}$ powder \cite{krot1985} pressed into the form of a pellet was investigated. With this type of target, carbon clusters C$_n$ with $n\geq 20$ have been produced (Fig.\,\ref{fig_c60}), which cover the mass range of the heavy elements to be investigated at \textsc{Triga-Trap}. 

The laser ion source at \textsc{Triga-Trap} will be used as well for the desorption and ionization of heavy elements in the transuranium region. The element of interest will be electrochemically deposited on a target plate. The production of heavy ions was tested with a gadolinium sample of $10^{11}$ atoms, which is the chemical homologue of curium.

\subsubsection{Expected performance}

The extraction efficiency of the gas jet from the target chamber to the on-line ion source has been previously determined to be between 60\,\% and 70\,\% \cite{stend1980}. The efficiency for surface ionization is about 6\,\%. The transport efficiency in
the trap section is estimated to be 80\,\%. The MCP detectors for the time-of-flight mass measurement detect about one third of the incoming ions and the channeltron reaches $>$\,90\,\%. \textsc{Triga-Trap} will therefore have a total efficiency of about 1\,\%, similar to existing facilities \cite{blau2003b}.

With the mass-selective buffer-gas cooling in the purification trap, a resolving power of typically $10^5$ can be obtained \cite{blau2004,webe2005a,sava1991,boll2001,roos2004}. The resolving power of the precision trap is given by the Fourier limit \cite{boll2001}
\begin{eqnarray}
\Re=\frac{m}{\Delta m}=\frac{\nu_c}{\Delta\nu_c}\approx\nu_cT,
\end{eqnarray}
where $m$ is the mass, $\nu_c$ the corresponding cyclotron frequency, and $\Delta\nu_c$ is the line width (\textsc{Fwhm}) of the frequency signal. $T$ is the excitation time in a time-of-flight measurement and in a \textsc{Ft-Icr} measurement the observation time.

The resolving power in a time-of-flight mass measurement is strongly depending on the half-life of the nuclide under investigation, since it limits the excitation time. With an excitation period of 1\,s and a typical cyclotron frequency in the MHz range, $\Re=10^6$ can be reached with the precision trap. In case of a \textsc{Ft-Icr} measurement and a 1\,s observation time, a similar resolving power is reached.

The statistical error of the mass determination is given by \cite{blau2006} 
\begin{eqnarray}
\left(\frac{\delta m}{m}\right)\approx\frac{1}{\Re\sqrt{N_{\textnormal{tot}}}},
\end{eqnarray}
where $N_{\textnormal{tot}}$ denotes either the number of ions in a time-of-flight resonance or the number of signal transients in the \textsc{Ft-Icr} measurement. With 1\,000 ions in a time-of-flight resonance, the statistical uncertainty for rather long-lived species ($T_{1/2}\,\gtrsim\,1$\,s) with a cyclotron frequency in the MHz range is about $3\cdot 10^{-8}$. When recording the image current signal with the non-destructive detection technique, the number of repeated measurement cycles will be strongly limited by the low production rates of the heavy nuclides. In this case, an accuracy of $10^{-6}$ to $10^{-7}$ is projected.

At this point it should be mentioned that not all of the previously discussed limits can be reached simultaneously, since the half-life of the nuclide of interest and its production rate are the determining factors. For half-lives longer than 1\,s and a sufficient production rate, the highest resolving power and accuracy can be obtained.

\subsection{The laser branch}

\subsubsection{Off-line ion source}

For the collinear laser-spectroscopic beamline two types of off-line ion sources will be installed. A commercial \textsc{Riber} Model \textsc{Ci-10 n6119} inert-gas electron impact source, typically applied for sputter gun applications, will be used for ion optical optimization. It provides high-flux ion beams with energies between 0.1\,-\,6\,keV. In first tests, intensities of up to 1\,nA of singly charged positive argon ions have been achieved. This is more than sufficient to optimize the ion beam profile and to develop the ion beam diagnostic tools. For first spectroscopic experiments, an electrothermally heated ion source will be used, which can efficiently provide surface ions of alkaline, some alkaline earth metals and lanthanides. Especially caesium and rubidium ions provide the possibility of precision tests of the absolute frequency determination and the high-precision voltage dividers after neutralization in the charge-exchange cell, since their resonance transitions have been determined to high accuracy \cite{ye1996,udem2000}. This ion source will be operated on a high-voltage platform to provide fast ion beams.

\subsubsection{Ion and laser beam diagnostics}

\label{sec_iondetection_laser} A combination of ion beam analysis with optical detection techniques will be employed for beam diagnostics in the \textsc{Triga-Laser} branch. It is mandatory to have appropriate tools to ensure good alignment of ion and laser beams for two reasons: only those ions, that are overlapped with the laser beam, contribute to the spectroscopic signal and any tilt between the axes of laser and ion beam causes systematic errors in the analysis. To achieve a coarse alignment, two iris diaphragms are placed in the beamline. For ion beams of high intensities two commercial Danfysik fork scanners will be used, which are oscillating in mechanical resonance. One will be placed within the 10$^{\circ}$ deflector chamber and a second one in the beam monitoring section at the end of the collinear beamline. Thus, not only the position of the ion beam can be monitored, but also the beam profile can be optimized utilizing pairs of steerers and quadrupole lenses before and after the 10$^{\circ}$ deflector unit.

\begin{figure}[tbp]
\begin{center}
\includegraphics*[width=0.48\textwidth]{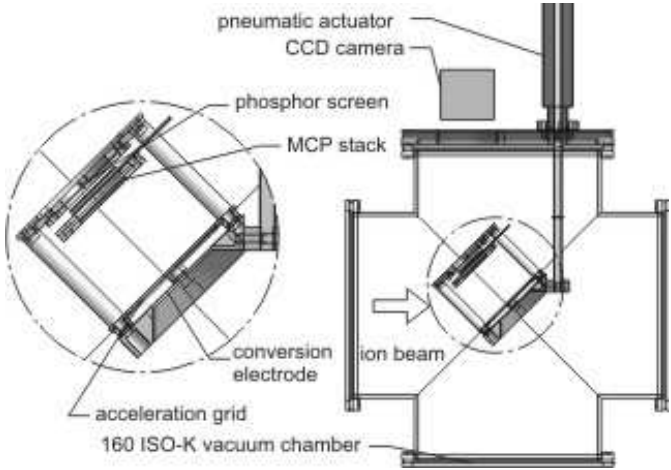}
\end{center}
\caption{Ion beam monitor for ion beams with intensities of a few thousand ions per second. The ion beam coming from the left hits a conversion electrode. The emitted electrons are accelerated by a grid on a variable potential, then amplified by a stack of MCPs and imaged on a phosphor screen, which is monitored by a CCD camera outside the vacuum. The complete module can be moved in and out of the beam by a pneumatic actuator.}
\label{fig_beammonitor}
\end{figure}

As radioactive ion beams of a few thousand ions per seconds do not generate sufficient current to be measured by fork scanners or in Faraday cups, dedicated devices for beam monitoring are required. A cut drawing of the \textsc{Triga-Laser} diagnostic device, similar to the one described in \cite{krug2000}, is shown in Fig.\,\ref{fig_beammonitor}. The compact detection system can be moved to the beam axis by pneumatic actuators. Coated metal plates serve as conversion electrodes which will be hit under 45$^{\circ}$ incidence angle by the ion beam. The emitted secondary electrons are accelerated by an overlaid grid on a stack of MCPs. A phosphorescence screen attached at the backside of the stack will be monitored by a CCD camera system. In this way, the spatial distribution of weak ion beams as well as their intensities can be monitored and adapted to the laser beam. The sensitivity of this detector type to ultra-violet and blue photons from the laser beam is presently not known. Therefore, a variety of vacuum compatible coatings for the conversion electrodes will be tested.

\subsubsection{Doppler tuning}

To bring the ions on resonance with the collinear laser beam, the velocity of the ions will be tuned by changing their kinetic energy. Since only a few ion and atomic species provide an energy level scheme with a closed two-level transition out of the ground-state, optical excitation is in most cases followed by decay into states that do not interact with the laser beam anymore. Thus, only a few photons will be scattered by the resonant ion or atom. To avoid this optical pumping process before the atom or ion reaches the optical detection region, interaction with the laser light must be suppressed. Therefore, the ion velocity is fine tuned by applying a potential to the charge-exchange cell (CEC) for spectroscopy of neutral atoms \cite{muel1983} or to a metal tube in the detection region for ion spectroscopy. While the ion source platform is kept at a fixed potential, the voltage of the CEC platform unit or the detection region can be scanned from -10\,kV to +10\,kV. As the linewidth and the accuracy of the transition frequency scale with the stability and accuracy of the potential, the CEC platform potential will be defined by a high-stability power supply providing a stability of better than $10^{-5}$ identical to the supply for the ion source.
For the determination of both platform potentials a new kind of high-voltage divider \cite{marx2001} will be set-up in collaboration with the group of Ch. Weinheimer at the University of M\"{u}nster \cite{thum2007} and tested for collinear laser spectroscopic purposes. The aim of this development is to obtain a high-voltage division from 60\,kV to a few volts with an absolute accuracy of better than 10\,ppm. Even though this accuracy is usually not needed for the spectroscopy of heavy elements, it would improve measurements of light ions like, e.g., Na or Mg, where the uncertainty of the acceleration voltage of typically $10^{-4}$ limits the accuracy of the extracted nuclear charge radii.

\subsubsection{Ion charge exchange}

\label{sec:ion_beam_neutralization}
\begin{figure}[tbp]
\begin{center}
\includegraphics*[width=0.48\textwidth]{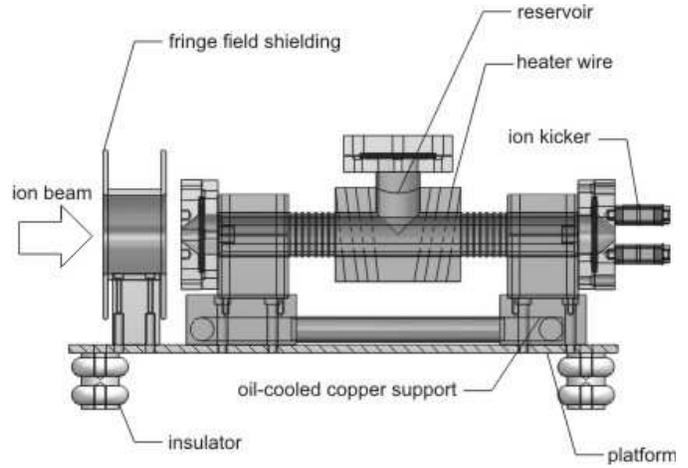}
\end{center}
\caption{Charge-exchange platform. A fringe field electrode minimizes the field gradient inside the charge-exchange cell, which is mounted on two oil-cooled copper blocks. These blocks serve as condensers for the alkaline vapor, which is generated by heating the reservoir. Residual ions are removed from the atom beam with an ion kicker at the exit of the charge-exchange device.}
\label{fig_dopplertuning}
\end{figure}

For the spectroscopy of fast atomic rather than ionic beams, the ions are neutralized in the charge-exchange cell \cite{baca1974} shown in Fig.\,\ref{fig_dopplertuning}. A reservoir in its center filled with an alkali metal can be kept at temperatures up to 600\,K by a wrapped coaxial thermo cable. Charge-exchange efficiencies above 90\,\% are expected at pressures in the regime of $10^{-2}$\,mbar \cite{verm1992}. Oil-cooled copper blocks at both ends of the CEC ensure the condensation of the alkali vapour and recirculation of the alkali metal inside the cell. Electrodes in front of the cell reduce the fringe fields at the entrance of the CEC to below 0.1\,V/mm. This avoids an ion-optical influence of the CEC potential on the beam. Furthermore, the conversion of an ion into an atom beam occurs in a region of constant potential. Residual ions not neutralized on the passage through the gas are deflected by a pair of electrostatic plates at the exit of the cell. 

\subsubsection{Fluorescence detection}

A purely optical detection is presently foreseen in the laser beamline. A section of a metal tube in sphero-ellipsoidal geometry which reflects the fluorescence light in a broad spectral region will focus the light efficiently via a lens system on an adjacent pair of photomultipliers \cite{mikh1969} or a row of channel photomultipliers. The latter can be operated in a wide wavelength range from about 200\,nm to 750\,nm with a quantum efficiency of up to 20\,\% and a dark count rate that is superior to most standard photomultipliers. The metal tube will be insulated against the vacuum chamber housing so that the tube can be biased for ion velocity tuning as described above. If collinear spectroscopy is performed on an atomic beam the optical detection region will be installed as close to the charge-exchange cell as possible to avoid optical pumping.

\subsubsection{Expected performance}

In the first version of collinear spectroscopy with the optical fluorescence detection a sensitivity limit of about $10^{5}$-$10^{6}$\,ions/s is expected. Even with this limited sensitivity a few interesting candidates can be studied, e.g., rhodium isotopes of which isotope shifts and hyperfine structures have not been measured so far. However, to extend isotope shift measurements further away from the valley of stability, more sensitive methods have to be adapted. It has been demonstrated at \textsc{Jyfl} that cooling and bunching of ion beams can clearly enhance the sensitivity and increase the signal-to-background ratio by up to four orders of magnitude  \cite{niem2002}. Using an RFQ cooler-buncher, laser resonance fluorescence signals have been obtained with ion yields as low as 100 ions/s and we expect a similar sensitivity once it will be added behind the mass separator at \textsc{Triga-Spec}.

Furthermore, resonance ionization spectroscopy (RIS) can be implemented into the \textsc{Triga-Laser} beamline. After preparation of a fast atomic beam, as it is described in Sec.\,\ref{sec:ion_beam_neutralization}, the atoms can be resonantly excited into an intermediate state which can then be efficiently ionized with pulses of an intense laser beam. Either a continuous wave (cw) or a pulsed laser can be applied for the resonance step offering high or medium resolution, respectively. The advantage of this method is that the created ions can be collected and detected with much higher efficiency and less background than the resonantly scattered photons, because fluorescence detection is always limited by the solid angle of the photon detector and the quantum efficiency of the photocathode. If RIS is being applied to a continuous atom beam, a high-repetition rate laser ($\nu_{rep}\approx 10\,$kHz) could be used for the ionization step \cite{schu1991}. When applied in combination with a cooler and buncher a low-repetition rate laser (10\,-\,100\,Hz) would be sufficient which has to be synchronized with the bunch that is being released from the RFQ cooler and buncher device.

A collinear RIS beamline has also recently been proposed for laser spectroscopy at the \textsc{Isolde} facility \cite{flan2008}. Generally, collinear RIS offers a very good signal-to-background ratio and a very high sensitivity and selectivity and has therefore also been applied for ultra-trace detection, e.g., of long-lived isotopes like $^{90}$Sr \cite{wend1997}. The main background stems from collisionally excited and/or ionized atoms or atoms in relatively high-lying metastable or long-lived high-angular momentum states that are created in the charge-exchange process. Those can often be ionized by the ionizing laser pulse without previous excitation with the resonant laser field and form therefore a constant, frequency-independent background that might limit the detection sensitivity. However, under favorable circumstances and good vacuum conditions, the sensitivity will probably be better than 100\,ions/s.

The ultimate sensitivity might be reached with ion trap experiments at the end of the \textsc{Triga-Laser} beamline. Ions could be decelerated with a pulsed drift tube and afterwards captured into an ion trap. The long observation times that are achievable for trapped ions make this technique superior to all other techniques if it comes to the investigation of species that have sufficiently long half-lives but are produced with extremely small rates, as it is the case for some transactinides and superheavy elements that are produced in fusion reactions. In-trap laser spectroscopy  and optical-microwave double resonance spectroscopy of ion species that have a complicated level structure and therefore unfavorable excitation schemes has been reported in Paul \cite{kael1989,beck1993} and Penning traps \cite{trap2003} to obtain hyperfine structure constants and nuclear $g$-factors. Buffer-gas cooled ion clouds have been used in these experiments to quench metastable states that would otherwise form dark states and render laser spectroscopy impossible due to the large number of repumping lasers that would be required. Even though we do not have definitive plans for such experiments yet, the \textsc{Triga-Laser} beamline is an excellent place to develop such methods using heavy actinides for, e.g., the spectroscopy of trans-actinides and superheavy elements that might later be applied at \textsc{Shiptrap}.

\subsubsection{Absolute transition frequencies and high-voltage determination in the collinear laser beam line}

The simultaneous measurement of transition frequencies in collinear ($\nu _{\mathrm{p}}$) and anti-collinear geometry ($\nu_{\mathrm{a}}$) allows us to determine the rest-frame frequency of the transition $\nu _{0}$ according to 
\begin{eqnarray}
\label{eq:AbsoluteTransFrequ}
\nu _{\mathrm{p}}\cdot \nu _{\mathrm{a}}=\nu _{0}^{2}.
\end{eqnarray}
Absolute transition frequencies can be determined with reasonable effort by application of a frequency comb, since fiber-laser-based combs are now sufficiently robust and reliable to be used in on-line applications. The comparison of the measured transition frequency, e.g.\,$\nu _{\mathrm{0}}$ in the collinear case, with the Doppler-shifted resonance frequency $\nu _{\mathrm{p}}=\nu _{0}\gamma (1+\beta )$ enables the determination of the beam velocity $\beta $ and, thus, the acceleration voltage $U_{\mathrm{ac}}$. Comparison with a direct voltage measurement using a sophisticated high-accuracy high-voltage divider will allow the determination of systematic uncertainties due to contact potentials and creeping currents.

An important point in this respect is the accurate alignment of the laser with the ion beams. The exact anti-parallel alignment of the collinear with the anti-collinear laser beams is of particular importance. Even small residual angles between the laser beams can cause relatively large errors in $\nu _{0}$. For example, a misalignment of 1\,mrad between the two laser beams will already induce an error of about 3\,MHz for a transition in the ultraviolett at $\lambda =313$\,nm of an ion with mass 10\,amu and 60\,kV acceleration voltage even though one of the laser beams might still be exactly parallel with the ion beam. In the mass region of calcium isotopes the situation is already less severe because the error under the same conditions is already reduced to about 1\,MHz. Small differences in the angle between the laser and the ion beam for different isotopes can therefore cause significant errors in the isotope shift. However, if the two laser beams are well aligned against each other and only the ion beam is not exactly parallel to the laser beams, the effect is reduced to about 10\,kHz for 1\,mrad misalignment. A precise antiprallel geometry can be realized by coupling the two lasers with fibers and making sure that the laser beam is being coupled into the fiber of the oppositely propagating laser beam after passage of the excitation region. A similar geometry has been used for tests of Special Relativity at the Test Storage Ring (TSR) in Heidelberg \cite{saat2003,rein2007}, where exact anti-collinear geometry was also required. The usability of such methods for the investigation of light elements will be explored at the \textsc{Triga-Laser} branch.

Transition frequencies in the $D1$-lines of alkali atoms are very accurately known and can serve as ideal candidates for such studies in different mass regions. Even though, the ultimate accuracy that might be reachable with this technique is not required for the isotope shift studies of the heavy and medium-mass elements available at the \textsc{Triga} reactor, these technical developments will be very useful for isotope shift studies on lighter elements at other facilities. These are ultimately limited by the small field shifts that are observable in the low-$Z$ region and the unavoidable systematic errors due to a limited accuracy of the acceleration voltage. Such investigations are currently performed for isotope shift measurements of beryllium isotopes planned at \textsc{Isolde} \cite{noer2008}.

\section{Project status and outlook}

The \textsc{Triga-Spec} facility offers the possibility to investigate neutron-rich nuclides, produced by neutron-induced fission in the research reactor \textsc{Triga} Mainz, as well as heavy transuranium elements. Currently, there is only one other facility planned for similar experiments, which is the \textsc{Caribu} project at the Argonne National Laboratory \cite{sava2006}. The combined mass spectrometric and laser spectroscopic setup in Mainz will provide reliable input data for calculations on the r-process in nucleosynthesis and for stringent tests of mass models in the heavy element region, as well as nuclear charge radii and nuclear moments from isotope shift and hyperfine structure measurements of neutron-rich nuclides. The technical developments in \textsc{Triga-Trap} allowing for single-ion detection and absolute mass measurements will have an impact on Penning trap mass spectrometry in the near future. For the first time, a detection system with a non-destructive single-ion sensitivity is applied at an on-line facility for mass measurements on short-lived nuclides, which provides the basis for experiments on rarely produced nuclides that have not been accessible up to now. \textsc{Triga-Trap} and \textsc{Triga-Laser} also serve as test facilities for mass and laser spectroscopic experiments at \textsc{Shiptrap} and the low-energy branch of the future \textsc{Fair} facility \cite{fair}.

The technical design of the \textsc{Triga-Spec} setup is finished and the construction almost completed. The trap branch assembly is completed and commissioning is ongoing. The major parts of the \textsc{Triga-Trap} setup like the ion transport optics, the Penning traps, and the image current detection system have been already tested. First off-line mass measurements on heavy elements with the laser ion source are planned to be performed in 2008. The heaviest nuclide which will be investigated in the Penning trap setup is \textsuperscript{252}Cf. In 2009, the setup of \textsc{Triga-Laser} as well as the on-line ion source and the mass separator magnet will be completed. The target chamber, which will be placed close to the reactor core, is currently being built. After all these installation works are done on-line mass and laser spectroscopic measurements on fission products at a nuclear reactor become possible.

\section*{Acknowledgements}

We acknowledge financial support by the Helmholtz Association for National Research Centers (\textsc{HGF}) under contracts VH--NG-037 and VH-NG-148. Szilard Nagy acknowledges financial support by the Humboldt Foundation. We thank H.-J. Kluge for his comments and his support while writing this paper. We also acknowledge the support from G. Hampel, C. Rauth, L. Schweikhard, and S. Stahl.




\begin{thebibliography}{99}

\bibitem{scha2006} H. Schatz, K. Blaum, Europhysics News 37 (5) (2006) 16.
\bibitem{burb1957} M. Burbidge et al., Rev. Mod. Phys. 29 (4) (1957) 548.
\bibitem{arno2003} M. Arnould, S. Goriely, Phys. Rep. 384 (2003) 1.
\bibitem{cowa2006} J.J. Cowan, C. Sneden, Nature 440 (2006) 1151.
\bibitem{blau2006} K. Blaum, Phys. Rep. 425 (2006) 1.
\bibitem{fair} See: http://www.gsi.de/fair/.
\bibitem{spir} See: http://ganinfo.in2p3.fr/research/developments/spiral2/index.html.
\bibitem{euri} See: http://www.ganil.fr/eurisol/.
\bibitem{ria} See: http://www.orau.org/RIA/.
\bibitem{eber2000} K. Eberhardt, A. Kronenberg, Kerntechnik 65 (2000) 5.
\bibitem{hamp2006} G. Hampel, K. Eberhardt, N. Trautmann, Atomwirtschaft 5 (2006) 328.
\bibitem{berk} See: http://ie.lbl.gov/fission.html.
\bibitem{habs2006} D. Habs et al., Eur. Phys. J. A 25 S01 (2005) 57.
\bibitem{sava2006} G. Savard et al., Int. J. Mass Spectrom. 251 (2006) 252.
\bibitem{stend1980} E. Stender, N. Trautmann, G. Herrmann, Radiochem. Radioanal. Lett. 42 (1980) 291.
\bibitem{raha2006} S. Rahaman et al., Int. J. Mass Spectrom. 251 (2006) 146.
\bibitem{joki2006} A. Jokinen et al., Int. J. Mass. Spectrom. 251 (2006) 204.
\bibitem{webe2005} C. Weber et al., Eur. Phys. J. A 25 S01 (2005) 25.
\bibitem{ferr2007} R. Ferrer et al., Eur. Phys. J. Special. Top. 150 (2007) 347.
\bibitem{koni1995} M. K\"{o}nig et al., Int. J. Mass. Spectrom. Ion. Proc. 142 (1995) 95.
\bibitem{mars1998} A.G. Marshall, C.L. Hendrickson, G.S. Jackson, Mass Spectrom. Revs. 17 (1998) 1.
\bibitem{yazi2007} C. Yazidjian et al., Hyp. Int. 173 (2006) 181.
\bibitem{kauf1976} S.L. Kaufmann, Opt. Comm. 17 (1976) 309.
\bibitem{anto1978} K.-R. Anton et al., Phys. Rev. Lett. 40 (1978) 642.
\bibitem{schi1978} B. Schinzler et al., Phys. Lett. B 79 (1978) 209.
\bibitem{klem1979} W. Klempt, J. Bonn and R. Neugart, Phys. Lett. B 82 (1979) 47.
\bibitem{wing1976} W.H. Wing et al., Phys. Rev. Lett. 39 (1976) 1488.
\bibitem{geit2000} W. Geithner et al., Hyp. Int. 129 (2000) 271.
\bibitem{niem2002} A. Nieminen et al., Phys. Rev. Lett.88 (2002) 094801.
\bibitem{neye2005} G. Neyens et al., Phys. Rev. Lett. 94 (2005) 022501.
\bibitem{neug2000} R. Neugart, Hyp. Int. 127 (2000) 101.
\bibitem{liev1992} P. Lievens et al., Nucl. Instr. Meth. B 70 (1992) 532.
\bibitem{noer2006} W. N\"{o}rtersh\"{a}user et al., Hyp. Int. 171 (2006) 149.
\bibitem{kubi2005} P. Kubina, Optics Express 13 (2005) 904.
\bibitem{sewt2003} M. Sewtz et al., Phys. Rev. Lett. 90 (2003) 163002.
\bibitem{back2007} H. Backe et al., Europ. Phys. J. D 45 (2007) 99.
\bibitem{brow1986} L.S. Brown, G. Gabrielse, Rev. Mod. Phys. 58 (1986) 233.
\bibitem{kell2003} A. Kellerbauer et al., Eur. Phys. J. D 22 (2003) 53.
\bibitem{grae1980} G. Gr\"{a}ff, H. Kalinowsky, J. Traut, Z. Phys. A 291 (1980) 35.
\bibitem{schw1991} L. Schweikhard, Int. J. Mass Spectrom. Ion Proc. 107 (1991) 281.
\bibitem{blau2004} K. Blaum et al., Europhys. Lett. 67 (2004) 586.
\bibitem{webe2005a} C. Weber et al., Phys. Lett. A 347 (2005) 81.
\bibitem {otte1989} E. W. Otten, \textit{Nuclear Radii and Moments of unstable Isotopes}. Treatise on Heavy Ion Science vol. 8: Nuclei far from stability, ed. Bromley, D.A. 1989, New York: Plenum Publishing Corp. (Springer) 515.
\bibitem{klug2003} H.-J. Kluge and W. N\"ortersh\"auser, Spectrochim. Acta B 58 (2003) 1031.
\bibitem{yan2000} Z.-C. Yan and G.W.F. Drake, Phys. Rev. A 61 (2000) 022504.
\bibitem{drak2005} G.W.F. Drake et al., Can. J. Phys. 83 (2005) 311.
\bibitem{hube1998} A. Huber et al., Phys. Rev. Lett. 80 (1998) 468.
\bibitem{muel2007} P. M\"{u}ller et al., Phys. Rev. Lett. 99 (2007) 252501.
\bibitem{sanc2006} R. S\'{a}nchez et al., Phys. Rev. Lett. 96 (2006) 033002.
\bibitem{alkh1983} G.D. Alkhazov et al., JETP Lett. 37 (1983) 274.
\bibitem{sauv2000} J. Sauvage et al., Hyp. Int. 129 (2000) 303.
\bibitem{bill1995} J. Billowes, P. Campbell, J. Phys. G 21 (1995) 707.
\bibitem{neug2002} R. Neugart, Eur. Phys. J. A 15 (2002) 35.
\bibitem{arno1987} E. Arnold et al., Phys. Lett. B 197 (1987) 311.
\bibitem{back1998} H. Backe et al., Phys. Rev. Lett. 80 (1998) 920.
\bibitem{herf2001} F. Herfurth et al., Nucl. Instrum. Meth. A 469 (2001) 254.
\bibitem{niem2001} A. Nieminen et al., Nucl. Instrum. Meth. A 469 (2001) 244.
\bibitem{mazu1980} A.K. Mazumar et al., Nucl. Instr. Meth. 174 (1980) 183.
\bibitem{menk1975} H. Menke, N. Trautmann, W.-J. Krebs, Kerntechnik 6 (1975) 281.
\bibitem{brue1985} M. Br\"{u}gger et al., Nucl. Inst. Meth. A 234 (1985) 218.
\bibitem{brue1979} M. Br\"{u}gger, Diploma thesis, University of Mainz (1979).
\bibitem{kugl2000} E. Kugler, Hyp. Int. 192 (2000) 23.
\bibitem{blau2003} K. Blaum et al., Eur. Phys. J. D 24 (2003) 145.
\bibitem{kell2002a} A. Kellerbauer et al., Hyp. Int. 146/147 (2003) 307.
\bibitem{chau2007} A. Chaudhuri et al., Eur. Phys. J. D 45 (2007) 47.
\bibitem{kirc1981} R. Kirchner et al., Nucl. Instr. Meth. 186 (1981) 295.
\bibitem{sim} See: http://www.simion.com.
\bibitem{beas1969} M.R. Beasley, R. Labusch, W.W. Webb, Phys. Rev. 181 (1969) 682.
\bibitem{sava1991} G. Savard et al., Phys. Lett. A 158 (1991) 247.
\bibitem{neid2007} D. Neidherr et al., Nucl. Instr. Meth. B, accepted (2008).
\bibitem{kret2008} M. Kretzschmar, submitted to Eur. Phys. J. A (2008).
\bibitem{gabr1983} G. Gabrielse, Phys. Rev. A 27 (1983) 2277.
\bibitem{webe2004} C. Weber, Dissertation, University of Heidelberg (2003), http://www.ub.uni-heidelberg.de/archiv/4435/.
\bibitem{wiza1979} J.L. Wiza, Nucl. Instr. Meth. 162 (1979) 587.
\bibitem{rain2004} S. Rainville, J.K. Thompson, D.E. Pritchard, Science 303 (2004) 334.
\bibitem{shi2005} W. Shi, M. Redshaw, E.G. Myers, Phys. Rev. A 72 (2005) 022510.
\bibitem{beck2004} D. Beck et al., Nucl. Instr. Meth. A 527 (2004) 567.
\bibitem{ring2006} R. Ringle et al., Int. J. Mass Spectrom. 251 (2006) 300.
\bibitem{blau2002} K. Blaum et al., Eur. Phys. J. A 15 (2002) 245.
\bibitem{blau2003a} K. Blaum et al., Analyt. Bioanalyt. Chem. 377 (2003) 1133.
\bibitem{krot1985} H.W. Kroto et al., Nature 318 (1985) 162.
\bibitem{blau2003b} K. Blaum et al., Phys. Rev. Lett. 91 (2003) 260801.
\bibitem{boll2001} G. Bollen, Nucl. Phys. A 693 (2001) 3.
\bibitem{roos2004} J. Van Roosbroeck et al., Phys. Rev. Lett. 92 (2004) 112501.
\bibitem{ye1996} J. Ye et al., Opt. Lett. 21 (1996) 1280.
\bibitem{udem2000} T. Udem et al., Phys. Rev. A (2000) 031801.
\bibitem{krug2000} K. Kruglov et al., Nucl. Instr. Meth. A 441 (2000) 595.
\bibitem{muel1983} A.C. Mueller et al., Nucl. Phys. A 403 (1983) 234.
\bibitem{marx2001} R. Marx, IEEE 50 (2001) 426.
\bibitem{thum2007} T. Th\"{u}mmler, Dissertation, University of M\"{u}nster (2007), http://miami.uni-muenster.de/servlets/DerivateServlet/Derivate-4212/diss\_thuemmler.pdf.
\bibitem{baca1974} M. Bacal and W. Reichelt, Rev. Sci. Instr. 45 (1974) 769.
\bibitem{verm1992} L. Vermeeren et al., J. Phys. B 25 (1992) 1009.
\bibitem{mikh1969} N.G. Mikhailus, J. Appl. Spectr. 11 (1969) 1123.
\bibitem{schu1991} Ch. Schulz et al., J. Phys. B 24 (1991) 4831.
\bibitem{flan2008} K. Flanagan et al., Proposal to the \textsc{Isolde} and \textsc{N-Tof} Committee, http://doc.cern.ch//archive/electronic/cern/preprints/intc/public/intc-2008-010.pdf.
\bibitem{wend1997} K. Wendt et al., Radiochim. Acta 79 (1997) 183.
\bibitem{kael1989} W. K\"alber et al., Z. Phys. A 334 (1989) 103.
\bibitem{beck1993} O. Becker et al., Phys.Rev.A 48 (1993) 3546.
\bibitem{trap2003} S. Trapp et al., Eur.Phys. J. D 26 (2003) 237.
\bibitem{saat2003} G. Saathoff et al., Phys. Rev. Lett. 91 (2003) 190403.
\bibitem{rein2007} S. Reinhardt et al., Nature Physics 3 (2007) 861.
\bibitem{noer2008} W. N\"ortersh\"auser et al., Proposal to the \textsc{Isolde} and \textsc{N-Tof} Committee, http://doc.cern.ch//archive/electronic/cern/preprints/intc/public/intc-2008-013.pdf.


\end{thebibliography}
\end{document}